# The spherical-harmonics representation for the interaction between diatomic molecules: the general case and applications to CO – CO and CO – HF.

Patricia R. P. Barreto [a], Ana Claudia P. S. Cruz [a], Rodrigo L. P. Barreto [b], Federico Palazzetti [c, *], Alessandra F. Albernaz [d], Andrea Lombardi [c], Glauciete S. Maciel [c, e], Vincenzo Aquilanti [c, f, g]

[a] Laboratório Associado de Plasma (LAP), Instituto Nacional de Pesquisas Espaciais (INPE)/MCT, CP 515, São José dos Campos, São Paulo, CEP 1224 7-9 70, Brazil.

[b] Universidade de Santa Catarina, Departamento de Engenharia Mecanica, 88040-900 Florianopolis, Santa Catarina – Brazil.

[c] Università di Perugia, Dipartimento di Chimica, Biologia e Biotecnologie, 06123 Perugia – Italy.

[d] Universidade de Brasìlia, Instituto de Fìsica – Campus Universitàrio Darcy Ribeiro, Brasìlia – Brazil.

[e] Current address: Secretària de Estado de Educação do Distrito Federal – Brasìlia - Brazil

[f] Consiglio Nazionale delle Ricerche, Istituto di Struttura della Materia, 00016 Roma, Italy

[g] Instituto de Física, Universidade Federal da Bahia, 40210 Salvador, Brazil

[*] Corresponding authors: federico.palazzetti@unipg.it.

Abstract

The spherical-harmonics expansion is a mathematically rigorous procedure and a powerful tool for the representation of potential energy surfaces of interacting molecular systems, determining their spectroscopic and dynamical properties, specifically in van der Waals clusters, with applications also to classical and quantum molecular dynamics simulations. The technique consists in the construction (by *ab initio* or semiempirical methods) of the expanded potential interaction up to terms that provide the generation of a number of leading configurations sufficient to account for faithful geometrical representations. This paper reports the full general description of the method of the spherical-harmonics expansion as applied to diatomic-molecule – diatomic-molecule systems of increasing complexity: the presentation of the mathematical background is given for providing both the application to the prototypical cases considered previously ($O_2$-$O_2$, $N_2$-$N_2$, and $N_2$-$O_2$ systems) and the generalization to: (i) the CO-CO system, where a characteristic feature is the lower symmetry order with respect to the cases studied before, requiring a larger number of expansion terms necessary to adequately represent the potential energy surface; and (ii) the CO – HF system, which exhibits the lowest order of symmetry among this class of aggregates and therefore the highest number of leading configurations.

# 1. Introduction.

In the last years, potential energy surfaces generated by spherical and hyperspherical harmonics expansions [1], have been successfully implemented in molecular dynamics simulations of non-reactive and reactive systems respectively [2-11]. These expansions are best suited for the calculation of matrix elements, which are needed in quantum mechanics. In classical molecular dynamics simulation, a specific requirement is a suitable representation of the interactions by a convenient analytical form, which permits a simple calculation of derivatives and a full account of the involved symmetries.

Multipolar expansions have been applied to several areas, including representation of potential energy surfaces, showing as a major feature the fast convergence of the series (see for example Van der Avoird *et al.* [12]). The multipolar expansion that we describe here is an exact transformation of quantum chemical (or experimental) input data related to a minimal number of configurations, called "leading configurations", selected on the basis of geometrical and physical characteristics of the system. Since the transformation is exact, the number of terms of the expansion corresponds to the number of leading configurations. The method permits interpolation and extrapolation as needed in dynamical and structure calculations. Spherical harmonics expansion has been largely used to characterize potential energy surfaces of a series of van der Waals aggregates using information from molecular beam studies and/or quantum chemical calculations: $H_2O$ – rare-gas-atom [13, 14], $N_2$ – $N_2$ [15], $O_2$ – $O_2$ [16], $N_2$ – $O_2$ [17], $N_2$ – NO [18], $H_2O$ – $H_2$ [19], $H_2O$ – $N_2$, $H_2O$ – $O_2$ [20], floppy-molecule – rare-gas-atom interactions [21-23]. Hyperspherical expansions have been successfully employed in molecular dynamics simulations for the study on oriented collisions of chiral molecules [24] and clusters dynamics [25, 26]. (For the state-of-art on potential energy surfaces of CO – CO dimers see [27], for van der Waals clusters, see for example [28-34]; for four-body systems see [35, 36]; for higher dimension systems see [37, 38] and references therein). The mathematical procedure consists in solving a finite dimensional linear algebra system, where the elements of the known vector are the input data, in this case the single energy points determined by quantum mechanical calculations, and the elements of the unknown vector are the expansion moments, thus the expansion in spherical harmonics of a certain configuration gives exactly the single point energy. The expansion moments provide interpolation among the potential curves corresponding to the leading configurations, to the whole configuration space. The key points of the method are expansibility by inclusion of further leading configurations and replacing of the input data, when more accurate ones are available.

In the characterization of the potential energy surface of a generic AB – CD van der Waals cluster (two diatomic molecules composed by four different atoms), we limit our considerations to non-reactive interactions and assume that the interatomic bonds are kept "frozen", *i. e.* the interacting molecules are considered as rigid, in their equilibrium position in the electronic ground state. The system is described by three vectors lying along the two A – B and C – D bonds, respectively, and a vector that joins the centers-of-mass of the two molecules. The interaction potential depends on four coordinates: the intermolecular distance $R$ between the centers-of-mass of the molecules; the $\theta_1$ and $\theta_2$ polar angles, and the dihedral angle $\phi$.

In the next section, we describe the method of representation of the potential energy surface: the coordinate system, the interaction potential and the leading configurations. In Section 3, we present an overview of the particular cases concerning binary interactions between diatomic molecules: systems with four identical atoms, we will review the $O_2$ - $O_2$ and $N_2$ - $N_2$ systems [15, 16], previously studied; with two different atoms: $N_2$ – $O_2$ [17], and CO – CO, for which we present detailed results; formulation of the expansion in spherical harmonics for systems of three different atoms, indicated by $A_2$ – BC, is also introduced. In Section 4, we present the formulation of the most general case, consisting of four different atoms, denoted as AB – CD. In Section 5, we show results on the application of this method on the CO – CO and CO – HF systems. Conclusions, in Section 6, close the paper.

## 2. Spherical harmonic expansion and representation of the potential energy surface.

_Coordinates._ The AB - CD system, where AB and CD are two diatomic molecules, is embedded in the Cartesian coordinate frame *xyz*, whose origin coincides with the center-of-mass of the whole system (in Figure 1, we report the coordinate reference system for CO – CO, but it can be extended to the general case AB - CD); the centers-of-mass of AB and CD are located on the *z*-axis and the *z'*- and *z''*- axes are parallel to the AB and CD bonds, respectively. The coordinates of the system are defined as follows: $R$ is the distance between the centers-of-mass of the two molecules, the polar angles $\theta_1$ and $\theta_2$ ($0 \leq \theta_1, \theta_2 \leq \pi$) are defined respectively as the angles between the *z'*- and the *z*-axis, and the *z''*- and the *z*-axis; the dihedral angle $\phi$ ($0 \leq \phi < 2\pi$) is given by the intersection between the semiplanes passing through the *z'*- and the *z*-axis, and the *z''*- and the *z*-axis.

The values of $\theta_1$ and $\theta_2$ are 0 and π when the A – B (and C – D) bonds are parallel to the *z*-axis. The definition of 0 and π is arbitrary, we set both angles equal to zero when A and D atoms are placed at values of the *z* coordinate lower than those of B and C atoms. The passage from $\theta_1$ (or $\theta_2$) = 0 to $\theta_1$ (or $\theta_2$) = π is allowed by an anti-clockwise rotation around the center-of-mass of the AB molecule and a clockwise rotation of the CD molecule. The dihedral angle $\phi$ is 0 and π when the semiplanes passing through the *z*- and *z'*- axes and the *z*- and *z''*- axes are coplanar; specifically, $\phi$ is 0 when A and C atoms (or B and D) have the same sign of the *y* coordinate. For values of $\theta_1$ or $\theta_2$ (or both) equal to zero, the dihedral angle $\phi$ is undetermined.

_Interaction potential._ The bond lengths of the molecules are considered "frozen", in such a way the interaction potential depends only on four variables: $R$, $\theta_1$, $\theta_2$, and $\phi$. The potential energy surface, $V$, is expanded into a series of appropriate angular functions, $F_{m,n}$ [1]:

$$V(R, \theta_1, \theta_2, \phi) = \sum_{m,n} v_{m,n}(R) F_{m,n}(\theta_1, \theta_2, \phi), \quad (1)$$

where $v_m(R)$ are the expansion moments and depend on the radial coordinate. In our case, the angular functions are bipolar spherical harmonics, $Y^{L0}_{L_1L_2(\theta_1,\theta_2,\phi)}$, where $L$, $L_1$ and $L_2$ ($|L_1 - L_2| \le L \le L_1 + L_2$) are non negative and integer numbers, and are elements of the Wigner 3-$j$ symbol

$$\begin{pmatrix} L_1 & L_2 & L \\ m & -m & 0 \end{pmatrix},$$

with $-\min(L_1, L_2) \le m \le \min(L_1, L_2)$, also integer. Thus, Eq. (1) becomes

$$V(R;\theta_1,\theta_2,\phi) = \sum_{L_1,L_2,m}\begin{pmatrix} L_1 & L_2 & L \\ m & -m & 0 \end{pmatrix} v_{L_1L_2L}(R)Y^m_{L_1}(\theta_1,0)Y^m_{L_2}(\theta_2,\phi). \quad (2)$$

*Leading configurations.* The leading configurations a specific choice of $\theta_1$, $\theta_2$, and $\phi$, made upon physical considerations on the geometrical features of the systems. The expansion moments $v_{m,n}(R)$ are the unknown values (see [1] and references therein) of a system of linear equations:

$$\begin{matrix} v_{11}(R)\,F_{11}(\theta_1,\theta_2,\phi) + & \cdots & +v_{1N}(R)F_{1N}(\theta_1,\theta_2,\phi) = V_1 \\ \vdots & \ddots & \vdots \\ v_{N1}(R)\,F_{N1}(\theta_1,\theta_2,\phi) + & \cdots & +v_{NN}(R)F_{NN}(\theta_1,\theta_2,\phi) = V_N \end{matrix} \quad (3)$$

where $V_1, \dots, V_N$ are the interaction potentials of the leading configurations, determined by *ab initio* or semiempirical methods. Each equation is the expansion of real spherical harmonics $F_{m,n}$, where $m$ denotes the spherical harmonics (common to all the equations) and $n$ is referred to the specific leading configuration. (The expansion moments and the spherical harmonics will be indicated in the paper as $v_{L_1L_2L}(R)$ and $Y^m_{L_1}(\theta_1,0)Y^m_{L_2}(\theta_2,\phi)$, respectively, as reported in Equation 2). Being the transformation of the input data exact, the interaction potentials calculated for the leading configuration correspond exactly to $V_1,\dots,V_N$. For the cases presented in this paper, the leading configurations can be divided into five classes: H, X, L, Z, and T, depending on the values of their angular variables. The H and Z configurations are characterized by a parallel displacement of the molecule: in the first case both $\theta$ are $\pi/2$, while in the second case $\theta$ is $\pi/4$; for both configurations, the dihedral angle $\phi$ is 0. In the X configuration, $\phi$ is $\pi/2$; while for T and L configurations $\phi$ is undetermined because of the linear displacement ($\theta = 0$) of one or both molecules, respectively.

In this work, we report an application of the method to the case of CO – CO and CO – HF systems. The potential energy surfaces have been generated by calculating the single point energies of 100 radial points for each leading configurat-ion. Calculations were performed for molecules in their most stable geometry. Both single point energies and optimizations were calculated at CCSD(T)/aug-cc-pVQZ level of theory.

## 3. Particular cases.

In this section, we report an overview of some case studies of diatomic molecule – diatomic molecule systems. We will indicate as $A_2$-$A_2$ the case of identical interacting homonuclear diatomic molecules, such as $N_2$-$N_2$ and $O_2$-$O_2$, and by $A_2$-$B_2$ that of different homonuclear diatomic molecules. The CO – CO system, which results are reported here, is denoted by AB – AB, while $A_2$-BC indicates the system composed by a heteronuclear and a homonuclear diatomic molecule, such as $N_2$ – NO. Finally, the general case, a system composed by four different atoms and indicated by AB – CD, is represented by CO – HF and the results are reported in this work.

### 3.1. Identical diatomic homonuclear molecules, the $A_2$ – $A_2$ case.

**3.1.a. The interaction potential expansion.** The $A_2$ – $A_2$ case has been addressed for the $N_2$ – $N_2$ and $O_2$ – $O_2$ systems [15, 16]. The potential energy surface can be adequately represented by five leading configurations: the *linear* L ($\theta_1 = \theta_2 = 0$); the *parallel* H ($\theta_1 = \theta_2 = \pi/2$, $\phi=0$); the *perpendicular* T ($\theta_1 = \pi/2$, $\theta_2 = 0$, $\phi= 0$); the tilted Z ($\theta_1 = \theta_2 = \pi/4$, $\phi=0$); finally, the only configuration, X, with a non-zero dihedral angle ($\theta_1 = \theta_2 = \pi/2$, $\phi = \pi/2$). Because of the system, equivalent $\theta$=0 and $\theta$=$\pi$, are equivalent. The interaction potential is given by

$$V_{A_2\ldots A_2}(R,\theta_1,\theta_2,\phi) = \Big[v_{000}(R) + \frac{\sqrt{5}}{4}v_{022}(R)\big(1 + 3\cos(2\theta_2)\big) + \frac{\sqrt{5}}{4}v_{202}(R)\big(1 + 3\cos(2\theta_1)\big) +$$

$$\frac{\sqrt{5}}{16}v_{220}(R)\big(1 + 3cos(2\theta_1)\big)\big(1 + 3cos(2\theta_2)\big) + 3\big(1 - cos(2\theta_1)\big)\big(1 - cos(2\theta_2)cos(\phi)\big) +$$

$$12cos(\phi)sin(2\theta_1)sin(2\theta_2) - \frac{5\sqrt{14}}{112}v_{222}(R)\big(1 + 3cos(2\theta_1)\big)\big(1 + 3cos(2\theta_2)\big) - 3\big(1 -$$

$$cos(2\theta_1)\big)\big(1 - cos(2\theta_2)cos(\phi)\big) + 6cos(\phi)sin(2\theta_1)sin(2\theta_2) + \frac{3\sqrt{70}}{224}v_{224}(R)\Big(2\big(1 +$$

$$3cos(2\theta_1)\big)\big(1 + 3cos(2\theta_2)\big) + \big(1 - cos(2\theta_1)\big)\big(1 - cos(2\theta_2)cos(2\phi)\big) -$$

$$16cos(\phi)sin(2\theta_1)sin(2\theta_2)\Big)\Big] \quad (4)$$

**3.1.b. The expansion moments.** The expansion moments are expressed in terms of the interaction potentials of the leading configurations. The isotropic term of the potential interaction is represented by the first term of the expansion moments $v_{000}(R)$, the other expansion moments represent the anisotropic terms.

(i) isotropic term:

$$v_{000}(R)_{A_2-A_2} = \frac{1}{9}\{2V_H(R) + V_L(R) + 2[2V_T(R) + V_X(R)]\}$$

(ii) anisotropic terms:

$$v_{022}(R)_{A_2-A_2} = \frac{1}{9\sqrt{5}}\{2V_H(R) - V_L(R) - V_T(R) + V_X(R)\}$$

$$v_{202}(R)_{A_2-A_2} = \frac{1}{9\sqrt{5}}\{2V_H(R) - V_L(R) + V_T(R) + V_X(R)\}$$

$$v_{220}(R)_{A_2-A_2} = \frac{1}{45\sqrt{5}}\{4V_H(R) - V_L(R) - 5[2V_T(R) + V_X(R)] + 12V_Z(R)\}$$

$$v_{222}(R)_{A_2-A_2} = \frac{1}{45}\sqrt{\frac{2}{7}}\{13V_H(R) - V_L(R) + 7[2V_T(R) - 2V_X(R)] - 12V_Z(R)\}$$

$$v_{224}(R)_{A_2-A_2} = \frac{8}{15}\sqrt{\frac{2}{7}}\{V_H(R) + V_L(R) + 2V_Z(R)\} \qquad (5)$$

### 3.2. Different diatomic homonuclear molecules, the $A_2 - B_2$ case.

**3.2.a. The interaction potential expansion.** The $N_2 - O_2$ system is a case study of $A_2 - B_2$ [17]. Compared to the previous case, it requires an additional leading configuration to adequately represent the potential energy surface. More precisely, it is necessary to distinguish, for T, between the configuration with $\theta_1 = \pi/2$ (related to the N – N bond) and $\theta_2 = 0$ (referred to the O – O bond), namely $T_1$, and the configuration with $\theta_1 = 0$ and $\theta_2 = \pi/2$, indicated as $T_2$. The interaction potential is expressed as follows

$$V_{A_2\ldots B_2}(R; \theta_1, \theta_2, \phi) = \Big[v_{000}(R) + \frac{\sqrt{5}}{4}v_{022}(R)\big(1 + 3cos(2\theta_2)\big) + \frac{\sqrt{5}}{4}v_{202}(R)\big(1 + 3cos(2\theta_1)\big) + \frac{\sqrt{5}}{16}v_{220}(R)\big(1 + 3cos(2\theta_1)\big)\big(1 + 3cos(2\theta_2)\big) + 3\big(1 - cos(2\theta_1)\big)\big(1 - cos(2\theta_2)cos(\phi)\big) + 12cos(\phi)sin(2\theta_1)sin(2\theta_2) - \frac{5\sqrt{14}}{112}v_{222}(R)\big(1 + 3cos(2\theta_1)\big)\big(1 + 3cos(2\theta_2)\big) - 3\big(1 - cos(2\theta_1)\big)\big(1 - cos(2\theta_2)cos(\phi)\big) + 6cos(\phi)sin(2\theta_1)sin(2\theta_2) + \frac{3\sqrt{70}}{224}v_{224}(R)\Big(2\big(1 + 3cos(2\theta_1)\big)\big(1 + 3cos(2\theta_2)\big) + \big(1 - cos(2\theta_1)\big)\big(1 - cos(2\theta_2)cos(2\phi)\big) - 16cos(\phi)sin(2\theta_1)sin(2\theta_2)\Big)\Big] \quad (6)$$

**3.2.b. The expansion moments.** The expansion moments are expressed in terms of the interaction potentials of the leading configurations

(i)      isotropic term:

$$v_{000}(R)_{A_2-B_2} = \frac{1}{9}\left\{2V_H(R) + V_L(R) + 2\left[V_{T_1}(R) + V_{T_2}(R) + V_X(R)\right]\right\}$$

(ii)     anisotropic term:

$$v_{022}(R)_{A_2-B_2} = \frac{1}{9\sqrt{5}}\left\{2V_H(R) - V_L(R) - 2V_{T_1}(R) + V_{T_2}(R) + V_X(R)\right\}$$

$$v_{202}(R)_{A_2-B_2} = \frac{1}{9\sqrt{5}}\left\{2V_H(R) - V_L(R) + V_{T_1}(R) - 2V_{T_2}(R) + V_X(R)\right\}$$

$$v_{220}(R)_{A_2-B_2} = \frac{1}{45\sqrt{5}}\left\{4V_H(R) - V_L(R) - 5\left[V_{T_1}(R) + V_{T_2}(R) + V_X(R)\right] + 12V_Z(R)\right\}$$

$$v_{222}(R)_{A_2-B_2} = \frac{1}{45}\sqrt{\frac{2}{7}}\left\{13V_H(R) - V_L(R) + 7\left[V_{T_1}(R) + V_{T_2}(R) - 2V_X(R)\right] - 12V_Z(R)\right\}$$

$$v_{224}(R)_{A_2-B_2} = \frac{8}{15}\sqrt{\frac{2}{7}}\left\{V_H(R) + V_L(R) + 2V_Z(R)\right\} \qquad (7)$$

## 3.3. Homonuclear – heteronuclear diatomic molecules: the $A_2$ – BC case.

**3.3.a. The interaction potential expansion.** Here, we report the case of a homonuclear and a heteronuclear interacting molecules, $A_2$ – BC, already applied to the system $N_2$ – NO [18]. The number of leading configurations is eight: two linear L, three perpendicular T, two tilted Z, a parallel H, and the X configuration. The interaction potential reads:

$$V_{A_2\ldots BC}(R; \theta_1, \theta_2, \phi)$$

$$= \Bigg[ v_{000}(R) + \sqrt{3}\, v_{011}(R)\big(\cos(\theta_2)\big) + \frac{\sqrt{5}}{4} v_{022}(R)\big(1 + 3\cos(2\theta_2)\big)$$

$$+ \frac{\sqrt{5}}{4} v_{202}(R)\big(1 + \cos(2\theta_1)\big) - \frac{1}{2}\frac{\sqrt{3}}{2} v_{211}(R)\big(1 + \cos(2\theta_1)\big)\cos(\theta_2)$$

$$- 3\cos(\phi)\sin(2\theta_1)\sin(\theta_2) + \frac{3}{4} v_{213}(R)\big(1 + \cos(2\theta_1)\big)\cos(\theta_2)$$

$$- 2\cos(\phi)\sin(2\theta_1)\sin(\theta_2) + \frac{\sqrt{5}}{16} v_{220}(R)\big(1 + \cos(2\theta_1)\big)\big(1 + \cos(2\theta_2)\big)$$

$$+ 3\big(1 - \cos(2\theta_1)\big)\big(1 - \cos(2\theta_2)\cos(\phi)\big) + 12\cos(\phi)\sin(2\theta_1)\sin(2\theta_2)$$

$$- \frac{5}{8\sqrt{14}} v_{222}(R)\big(1 + \cos(2\theta_1)\big)\big(1 + \cos(2\theta_2)\big)$$

$$+ 3\big(1 - \cos(2\theta_1)\big)\big(1 - \cos(2\theta_2)\cos(\phi)\big) - 6\cos(\phi)\sin(2\theta_1)\sin(2\theta_2)$$

$$+ \frac{3}{16}\sqrt{\frac{5}{14}} v_{224}(R)\Big(2\big(1 + \cos(2\theta_1)\big)\big(1 + \cos(2\theta_2)\big)$$

$$+ \big(1 - \cos(2\theta_1)\big)\big(1 - \cos(2\theta_2)\cos(2\phi)\big)$$

$$- 16\cos(\phi)\sin(2\theta_1)\sin(2\theta_2)\Big) \Bigg] \quad (8)$$

**3.3.b. The expansion moments.** The expansion moments are expressed in terms of the interaction potentials of the leading configurations:

(i)  isotropic component

$$v_{000}(R)_{A_2-BC} = \frac{1}{18}\Big\{4V_H(R) + V_{L_1}(R) + V_{L_2}(R)$$
$$+ 2\left[V_{T_1}(R) + V_{T_2}(R) + 2\left(V_{T_3}(R) + V_X(R)\right)\right]\Big\}$$

(ii) anisotropic component

$$v_{011}(R)_{A_2-BC} = \frac{1}{6\sqrt{3}}\Big\{V_{L_1}(R) - V_{L_2}(R) + 2V_{T_1}(R) - 2V_{T_2}(R)\Big\}$$

$$v_{211}(R)_{A_2-BC} = \frac{1}{30\sqrt{3}}\Big\{6V_H(R) + \left(3 - 2\sqrt{2}\right)V_{L_1}(R) + \left(3 + 2\sqrt{2}\right)V_{L_2}(R) + \left(3 + 2\sqrt{2}\right)V_{T_1}(R)$$
$$+ \left(3 - 2\sqrt{2}\right)V_{T_2}(R) + 6\left[V_{T_3}(R) - 2\left(V_{Z_1}(R) + V_{Z_2}(R)\right)\right]\Big\}$$

$$v_{213}(R)_{A_2-BC} = \frac{1}{30}\Big\{2\sqrt{2}V_H(R) + \left(2 + 2\sqrt{2}\right)V_{L_1}(R) + \left(-2 + \sqrt{2}\right)V_{L_2}(R) + \left(-2 + \sqrt{2}\right)V_{T_1}(R)$$
$$+ \left(2 + 2\sqrt{2}\right)V_{T_2}(R) + 2\sqrt{2}\left[V_{T_3}(R) - 2\left(V_{Z_1}(R) + V_{Z_2}(R)\right)\right]\Big\}$$

$$v_{022}(R)_{A_2-BC} = \frac{1}{9\sqrt{5}}\Big\{-2V_H(R) + V_{L_1}(R) + V_{L_2}(R) - 2\left[V_{T_1}(R) + V_{T_2}(R) - V_{T_3}(R) + V_X(R)\right]\Big\}$$

$$v_{202}(R)_{A_2-BC} = \frac{1}{9\sqrt{5}}\Big\{-2V_H(R) + V_{L_1}(R) + V_{L_2}(R) - V_{T_1}(R) - 2V_{T_2}(R) + 4V_{T_3}(R) + V_X(R)\Big\}$$

$$v_{220}(R)_{A_2-BC} = \frac{1}{45\sqrt{5}}\Big\{14V_H(R) + \left(2 - 3\sqrt{2}\right)V_{L_1}(R) + \left(2 + 3\sqrt{2}\right)V_{L_2}(R) - 5(2 +$$
$$3\sqrt{2})\left[V_{T_1}(R) + \left(-2 + 3\sqrt{2}\right)V_{T_2}(R) - 2[2V_{T_3}(R)] - 6\left(V_{Z_1}(R) - V_{Z_2}(R)\right)\right]\Big\}$$

$$v_{222}(R)_{A_2-BC} = \frac{1}{45}\sqrt{7}\Big\{10\sqrt{2}V_H(R) + \big(3+2\sqrt{2}\big)V_{L_1}(R) - \big(3-2\sqrt{2}\big)V_{L_2}(R)$$

$$+ \big(3+2\sqrt{2}\big)\Big[V_{T_1}(R) + \big(3+2\sqrt{2}\big)V_{T_2}(R) + 2\sqrt{2}\big[2V_{T_3}(R) - 7V_X(R)\big]$$

$$- 3\Big(V_{Z_1}(R) - V_{Z_2}(R)\Big)\Big]\Big\}$$

$$v_{224}(R)_{A_2-B} = \frac{2}{15}\sqrt{\frac{2}{35}}\Big\{2V_H(R) + \big(1+\sqrt{2}\big)V_{L_1}(R) + \big(1+\sqrt{2}\big)V_{L_2}(R) - \sqrt{2}V_{L_3}(R) + \big(-1+$$

$$\sqrt{2}\big)V_{T_1}(R) - \big(1+\sqrt{2}\big)V_{T_2}(R) - 2\big[V_{T_3}(R) + 2V_{Z_1}(R) - 2V_{Z_2}(R)\big]\Big\} \qquad (9)$$

### 3.4. Heteronuclear identical diatomic molecules: the AB – AB case.

In this work, we present the CO – CO system as a case study of AB – AB. In Figure 2, we show the eleven leading configurations used to represent the potential energy surface: with respect the cases discussed in the two previous subsections, the presence of heteronuclear molecules requires a higher number of leading configurations, to adequately describe the potential energy surface. A right choice of them requires three L configurations, two H, three Z, two T and one X.

The explicit form of the interaction potential is

$$V_{AB-AB}(R,\theta_1,\theta_2,\phi) = v_{000}(R) + \sqrt{3}v_{101}(R)(\cos\theta_1 + \cos\theta_2) -$$

$$\sqrt{3}v_{110}(R)(\sin\theta_1\sin\theta_2\cos\phi + \cos\theta_1\cos\theta_2) +$$

$$v_{112}(R)\left(\sqrt{6}\cos\theta_1\cos\theta_2 - \sqrt{\frac{3}{2}}\sin\theta_1\sin\theta_2\cos\phi\right) -$$

$$v_{121}(R)\frac{1}{2}\sqrt{\frac{3}{2}}\big(3\sin\theta_1\sin(2\theta_2)cos\phi + \cos\theta_1\left(3\cos(2\theta_2) + 1\right)\big) +$$

$$v_{123}(R)\left(\frac{3}{4}\cos\theta_1(3\cos(2\theta_2) + 1) - \frac{3}{2}\sin\theta_1\sin(2\theta_2)\cos\phi\right) +$$

$$\frac{3\sqrt{5}}{4}v_{202}(R)(\cos(2\theta_1) + \cos(2\theta_1)) - v_{211}(R)\left(\frac{3}{2}\sqrt{\frac{3}{2}}\sin(2\theta_1)\sin(\theta_2)\cos\phi +\right.$$

$$\frac{1}{2}\sqrt{\frac{3}{2}}\left(3\cos(2\theta_1)+1\right)\cos\theta_2 + v_{213}(R)\left(\frac{3}{4}(3\cos(2\theta_1)+1)\cos(\theta_2)-\right.$$

$$\left.\frac{3}{2}\sin(2\theta_1)\sin\theta_2\cos\phi\right) + v_{220}(R)\left(\frac{3\sqrt{5}}{16}(1-\cos(2\theta_1))(1-\cos(2\theta_2))\cos(2\phi)+\right.$$

$$\left.\frac{3\sqrt{5}}{4}\sin(2\theta_1)\sin(2\theta_2)\cos\phi + \frac{\sqrt{5}}{16}(3\cos(2\theta_1)+1)(3\cos(2\theta_2)+1)\right)-$$

$$v_{222}(R)\left(-\frac{15(1-\cos(2\theta_1))(1-\cos(2\theta_2))\cos(2\phi)}{8\sqrt{14}}+\frac{15\sin(2\theta_1)\sin(2\theta_2)\cos\phi}{4\sqrt{14}}+\right.$$

$$\left.\frac{5(3\cos(2\theta_1)+1)(3\cos(2\theta_2)+1)}{8\sqrt{14}}\right) + v_{224}(R)\left(\frac{3}{16}\sqrt{\frac{5}{14}}(1-\cos(2\theta_1))(1-\cos(2\theta_2))\cos(2\phi)-\right.$$

$$\left.3\sqrt{\frac{5}{14}}\sin(2\theta_1)\sin(2\theta_2)\cos\phi + \frac{3}{8}\sqrt{\frac{5}{14}}(3\cos(2\theta_1)+1)(3\cos(2\theta_2)+1)\right) \quad (10)$$

The expansion moments expressed in terms of the interaction potentials of the leading configurations are

(i) isotropic term:

$$v_{000}(R)_{AB-AB}=\frac{1}{36}\left(4\left(V_{H_1}(R)+V_{H_2}(R)\right)+\left(V_{L_1}(R)+2V_{L_2}(R)+V_{L_3}(R)\right)+8\left(V_{T_1}(R)+V_{T_2}(R)\right)\right.$$
$$\left.+8V_X(R)\right)$$

(ii) anisotropic terms:

$$v_{110}(R)_{AB-AB}=-\frac{1}{12\sqrt{3}}\left(4\left(V_{H_1}(R)-V_{H_2}(R)\right)+\left(V_{L_1}(R)-2V_{L_2}(R)+V_{L_3}(R)\right)\right)$$

$$v_{112}(R)_{AB-A}=\frac{1}{6\sqrt{6}}\left(2\left(-V_{H_1}(R)+V_{H_2}(R)\right)+\left(V_{L_1}(R)-2V_{L_2}(R)+V_{L_3}(R)\right)\right)$$

$$v_{101}(R)_{AB-AB}=\frac{1}{12\sqrt{3}}\left(\left(V_{L_1}(R)-V_{L_3}(R)\right)+4\left(V_{T_1}(R)-V_{T_2}(R)\right)\right)$$

$$v_{121}(R)_{AB-AB}=\frac{1}{60\sqrt{3}}\left(12\left(-V_{H_1}(R)+V_{H_2}(R)\right)+2\sqrt{2}\left(-V_{T_1}(R)+V_{T_2}(R)\right)+24\left(V_{Z_2}(R)-V_{Z_3}(R)\right)\right.$$
$$\left.+\left((6-5\sqrt{2})V_{L_1}(R)-12V_{L_2}(R)+\left(6+5\sqrt{2}\right)V_{L_3}(R)\right)\right)$$

$$v_{123}(R)_{AB-AB} = \frac{1}{15\sqrt{2}}\Big(2\left(-V_{H_1}(R) + V_{H_2}(R)\right) + \left(V_{L_1}(R) - 2V_{L_2}(R) + V_{L_3}(R)\right)$$
$$+ 2\sqrt{2}\left(-V_{T_1}(R) + V_{T_2}(R)\right) + 4\left(V_{Z_2}(R) - V_{Z_3}(R)\right)\Big)$$

$$v_{202}(R)_{AB-AB} = \frac{1}{18\sqrt{5}}\Big(-2\left(V_{H_1}(R) + V_{H_2}(R)\right) + \left(V_{L_1}(R) + 2V_{L_2}(R) + V_{L_3}(R)\right) + 2\left(V_{T_1}(R) + V_{T_2}(R)\right)$$
$$- 4V_X(R)\Big)$$

$$v_{211}(R)_{AB-AB} = \frac{1}{60\sqrt{3}}\Big(12\left(V_{H_1}(R) - V_{H_2}(R)\right) + 2\sqrt{2}\left(-V_{T_1}(R) + V_{T_2}(R)\right) + 24\left(V_{Z_2}(R) - V_{Z_3}(R)\right)$$
$$- \left((6 + 5\sqrt{2})V_{L_1}(R) + 12V_{L_2}(R) + (-6 + 5\sqrt{2})V_{L_3}(R)\right)\Big)$$

$$v_{213}(R)_{AB-AB} = \frac{1}{30}\Big(2\sqrt{2}\left(V_{H_1}(R) - V_{H_2}(R)\right) + \sqrt{2}\left(-V_{L_1}(R) + 2V_{L_2}(R) - V_{L_3}(R)\right)$$
$$+ 4\left(-V_{T_1}(R) + V_{T_2}(R)\right) + \sqrt{2}\left(V_{Z_1}(R) - V_{Z_2}(R)\right)\Big)$$

$$v_{220}(R)_{AB-A} = \frac{1}{90\sqrt{5}}\Big(8\left(V_{H_1}(R) + 4V_{H_2}(R)\right) + \left(11V_{L_1}(R) - 2V_{L_2}(R) + 11V_{L_3}(R)\right)$$
$$+ 4\left(V_{T_1}(R) + V_{T_2}(R)\right) - 20V_X(R) - 24\left(V_{Z_1}(R) + V_{Z_2}(R)\right)\Big)$$

$$v_{222}(R)_{AB-A} = \frac{1}{90\sqrt{14}}\Big(2\left(13V_{H_1}(R) + V_{H_2}(R)\right) + \left(13V_{L_1}(R) - 2V_{L_2}(R) - 13V_{L_3}(R)\right)$$
$$+ 4\left(V_{T_1}(R) + V_{T_2}(R)\right) - 56V_X(R) + 24\left(V_{Z_1}(R) + V_{Z_3}(R)\right)\Big)$$

$$v_{224}(R)_{AB-AB} = -\frac{2}{15}\sqrt{\frac{2}{35}}\Big(2\left(-V_{H_1}(R) + V_{H_2}(R)\right) + \left(V_{L_1}(R) - 2V_{L_2}(R) + V_{L_3}(R)\right) + 4\left(V_{T_1}(R) + V_{T_2}(R)\right) - 4\left(V_{Z_1}(R) + V_{Z_3}(R)\right)\Big) \quad (11)$$

## 4. General case.

The systems illustrated in this section are particular cases of a general one, which considers four different atoms, AB – CD. Absence of symmetry relations makes this system the most complex of those discussed so far and the number of leading configurations required to build the potential energy surface is fourteen. The linear configurations L are in this case four as far as the perpendicular T, there are thus two parallel configurations H, three tilted configurations Z, and the X configuration (Figure 3).

## 4.1. Formulation.

The interaction potential obtained as expansion on fourteen moments is

$$V_{AB-C}\ (R, \theta_1, \theta_2, \phi) = v_{000}(R) + \sqrt{3}v_{011}(R)\cos\theta_2 + \frac{\sqrt{5}}{4}v_{022}(R)(3\cos(2\theta_2) + 1) +$$

$$\sqrt{3}v_{101}(R)\cos\theta_1 - v_{110}(R)\big(\sqrt{3}\sin\theta_1\sin\theta_2\cos\phi + \sqrt{3}\cos\theta_1\cos\theta_2\big) +$$

$$v_{112}(R)\left(\sqrt{6}\cos\theta_1\cos\theta_2 - \sqrt{\frac{3}{2}}\sin\theta_1\sin\theta_2\cos\phi\right) - v_{121}(R)\left(\frac{3}{2}\sqrt{\frac{3}{2}}\sin\theta_1\sin(2\theta_2\cos\phi) + \right.$$

$$\left. \frac{1}{2}\sqrt{\frac{3}{2}}\cos\theta_1(3\cos(2\theta_2) + 1)\right) + v_{123}(R)\left(\frac{3}{4}\cos\theta_1(3\cos(2\theta_2) + 1) - \frac{3}{2}\sin\theta_1\sin(2\theta_2)\cos\phi\right) +$$

$$\frac{\sqrt{5}}{4}v_{202}(R)(3\cos(2\theta_1) + 1) - v_{211}(R)\left(\frac{3}{2}\sqrt{\frac{3}{2}}\sin(2\theta_1)\sin(\theta_2)\cos\phi + \frac{1}{2}\sqrt{\frac{3}{2}}(3\cos(2\theta_1) + \right.$$

$$\left. 1)\cos\theta_2\right) + v_{213}(R)\left(\frac{3}{4}(3\cos(2\theta_1) + 1)\cos(\theta_2) - \frac{3}{2}\sin(2\theta_1)\sin\theta_2\cos\phi\right) + v_{220}(R)\left(\frac{3\sqrt{5}}{16}(1 - \right.$$

$$\left. \cos(2\theta_1))(1 - \cos(2\theta_2))\cos(2\phi) + \frac{3\sqrt{5}}{4}\sin(2\theta_1)\sin(2\theta_2)\cos\phi + \frac{\sqrt{5}}{16}(3\cos(2\theta_1) + \right.$$

$$\left. 1)(3\cos(2\theta_2) + 1)\right) - v_{222}(R)\left(-\frac{15(1 - \cos(2\theta_1))(1 - \cos(2\theta_2))\cos(2\phi)}{8\sqrt{14}} + \frac{15\sin(2\theta_1)\sin(2\theta_2)\cos\phi}{4\sqrt{14}} + \right.$$

$$\left. \frac{5(3\cos(2\theta_1) + 1)(3\cos(2\theta_2) + 1)}{4\sqrt{14}}\right) + v_{224}(R)\left(\frac{3}{16}\sqrt{\frac{5}{14}}(1 - \cos(2\theta_1))(1 - \cos(2\theta_2))\cos(2\phi) - \right.$$

$$\left. 3\sqrt{\frac{5}{14}}\sin(2\theta_1)\sin(2\theta_2)\cos\phi + \frac{3}{8}\sqrt{\frac{5}{14}}(3\cos(2\theta_1) + 1)(3\cos(2\theta_2) + 1)\right) \quad (12)$$

## 4.2. Expansion moments.

The expansion moments expressed in terms of the interaction potentials of the leading configurations are reported as follows:

(i) isotropic term

$$v_{000}(R)_{AB-CD} = \frac{1}{36}\Big[4\big(V_{H_1}(R) + V_{H_2}(R) + V_{T_1}(R) + V_{T_2}(R) + V_{T_3}(R) + V_{T_4}(R)\big) + V_{L_1}(R)$$

$$+ V_{L_2}(R) + V_{L_3}(R) + V_{L_4}(R) + 8V_X(R)\Big]$$

(ii) anisotropic terms

$$v_{011}(R)_{AB-} = \frac{V_{L_1}(R) - V_{L_2}(R) + V_{L_3}(R) - V_{L_4}(R) - 4\big(V_{T_1}(R) + V_{T_2}(R)\big)}{12\sqrt{3}}$$

$$v_{022}(R)_{AB-CD} = \frac{1}{18\sqrt{5}}\Big[2\big(-V_{H_1}(R) - V_{H_2}(R) - V_{T_3}(R) - V_{T_4}(R)\big) + 4\big(V_{T_1}(R) + V_{T_2}(R) - V_X(R)\big)$$

$$+ V_{L_1}(R) + V_{L_2}(R) + V_{L_3}(R) + V_{L_4}(R)\Big]$$

$$v_{110}(R)_{AB-CD} = \frac{4\big(-V_{H_1}(R) + V_{H_2}(R)\big) - V_{L_1}(R) + V_{L_2}(R) + V_{L_3}(R) - V_{L_4}(R)}{12\sqrt{3}}$$

$$v_{112}(R)_{AB-CD} = \frac{2\big(-V_{H_1}(R) + V_{H_2}(R)\big) + V_{L_1}(R) - V_{L_2}(R) - V_{L_3}(R) + V_{L_4}(R)}{6\sqrt{6}}$$

$$v_{101}(R)_{AB-C} = \frac{V_{L_1}(R) + V_{L_2}(R) - V_{L_3}(R) - V_{L_4}(R) + 4\big(V_{T_3}(R) - V_{T_4}(R)\big)}{12\sqrt{3}}$$

$$v_{121}(R)_{AB-CD} = -\frac{1}{60\sqrt{3}}\Big(6\big(3V_{H_1}(R) - V_{H_2}(R)\big) + \sqrt{2}\big(-V_{L_1}(R) + 5V_{L_2}(R) - 5V_{L_3}(R) + V_{L_4}(R)\big)$$

$$+ \big(6 - 6\sqrt{2}\big)V_{T_1}(R) + \big(6 + 6\sqrt{2}\big)V_{T_2}(R) + \big(6 - 4\sqrt{2}\big)V_{T_3}(R) + \big(6 + 6\sqrt{2}\big)V_{T_4}(R)$$

$$- 24\big(V_{Z_2}(R) + V_{Z_3}(R)\big)\Big)$$

$$v_{123}(R)_{AB-CD} = \frac{1}{60}\Big(2\sqrt{2}\big(-3V_{H_1}(R) + V_{H_2}(R)\big) + \big(4-\sqrt{2}\big)V_{L_1}(R) - \sqrt{2}V_{L_2}(R) - \sqrt{2}V_{L_3}(R)$$
$$- \big(4+\sqrt{2}\big)V_{L_4}(R) + 2V_{T_1}(R) - 6V_{T_2}(R) - 6V_{T_3}(R) + 2V_{T_4}(R)$$
$$+ 8\big(V_{Z_2}(R) + V_{Z_3}(R)\big)\Big)$$

$$v_{202}(R)_{AB-CD} = \frac{1}{18\sqrt{5}}\Big(2\big(-V_{H_1}(R) - V_{H_2}(R) - V_{T_1}(R) - V_{T_2}(R) + V_{L_1}(R) + V_{L_2}(R) + V_{L_3}(R)$$
$$+ V_{L_4}(R)\big) + 4\big(V_{T_3}(R) + V_{T_4}(R) - V_X(R)\big)\Big)$$

$$v_{211}(R)_{AB-} = \frac{1}{60\sqrt{3}}\Big(\big(3+\sqrt{2}\big)V_{L_1}(R) + \big(3+5\sqrt{2}\big)V_{L_2}(R) + \big(3-5\sqrt{2}\big)V_{L_3}(R)$$
$$+ \big(3-\sqrt{2}\big)V_{L_4}(R) + \big(6+4\sqrt{2}\big)V_{T_1}(R) + \big(6-4\sqrt{2}\big)V_{T_2}(R)$$
$$+ \big(6+6\sqrt{2}\big)V_{T_3}(R)\big(6-6\sqrt{2}\big)V_{T_4}(R)\big) - 24\big(V_{Z_1}(R) + V_{Z_2}(R)\big)$$

$$v_{213}(R)_{AB-C} = \frac{1}{60}\sqrt{2}\Big(6V_{H_1}(R) - 2V_{H_2}(R) + 5V_{L_1}(R) + V_{L_2}(R) + V_{L_3}(R) - 3V_{L_4}(R) - 2V_{T_1}(R)$$
$$+ 6V_{T_2}(R) + 6V_{T_3}(R) - 2V_{T_4}(R) - 8\big(V_{Z_1}(R) + 8V_{Z_2}(R)\big)\Big)$$

$$v_{220}(R)_{AB-CD} = \frac{1}{90\sqrt{5}}\Big(4V_{H_1}(R) + 20\big(-V_{H_2}(R) + V_X(R)\big) + 7\big(V_{L_1}(R) + V_{L_4}(R)\big)$$
$$+ 5\big(-V_{L_2}(R) - V_{L_3}(R)\big) + 10\big(V_{T_1}(R) + V_{T_2}(R) + V_{T_3}(R) + V_{T_4}(R)\big)$$
$$+ 24\big(-V_{Z_1}(R) - V_{Z_3}(R)\big)\Big)$$

$$v_{222}(R)_{AB-DE} = \frac{1}{90\sqrt{14}} \Big( 38 V_{H_1}(R) + 5 \left( V_{L_1}(R) + V_{L_4}(R) \right) + 7 \left( V_{L_2}(R) - V_{L_3}(R) \right)$$

$$+ 14 \left( V_{H_2}(R) + V_{T_1}(R) + V_{T_2}(R) + V_{T_3}(R) + V_{T_4}(R) - 56 V_X(R) \right)$$

$$+ 24 \left( V_{Z_1}(R) - V_{Z_3}(R) \right) \Big)$$

$$v_{224}(R)_{AB-D} = \frac{4}{15} \sqrt{\frac{2}{35}} \Big( 2 V_{H_1}(R) + V_{L_1}(R) + V_{L_4}(R) - 2 V_{Z_1}(R) - 2 V_{Z_3}(R) \Big) \qquad (13)$$

## 5. Applications.

Here, we present the results related to the calculation of the potential energy surfaces of CO – CO and CO – HF systems. We report the energy profiles as a function of the intermolecular distance for each leading configuration and then show the more significant *ab initio* points, *i. e.* those corresponding to the configurations closest to the minima, interpolated through the interaction potential reported in Equations (10) and (12). The *ab initio* calculations were carried out by using the MOLPRO [39] software at CCSD(T)/aug-cc-pVQZ level of theory. For each leading configuration, we determine 100 single energy points.

### 5.1. The CO – CO case.

Figure 4 reports the potential energy profiles of the leading configurations of the CO – CO system. In order to make the figure more readable, the plots have been displaced in four different panels. The most stable configurations are $T_1$ and $Z_1$, whose behavior is very similar: the minimum energy is *ca.* -120 cm$^{-1}$ and the equilibrium distance is *ca.* 4.2 Å; Z presents also two other stable configurations at *ca.* 4 Å: $Z_2$, which equilibrium energy is -110 cm$^{-1}$ and $Z_3$, at a slightly higher energy, *ca.* -100 cm$^{-1}$; still at 4 Å, there is another the $T_2$ configuration, which minimum is *ca.* -120 cm$^{-1}$. The equilibrium distance of the X configuration is located at *ca.* 3.7 Å and *ca.* -90 cm$^{-1}$; at similar values of energy and distance is also the minimum of $H_2$, while $H_1$ is less stable, *ca.* -50 cm$^{-1}$ at 4 Å. The least stable configurations are Ls, as expected, because of their linear reciprocal orientation: $L_3$ presents an equilibrium energy of *ca.* -40 cm$^{-1}$ at 4.5 Å, $L_2$ has a minimum at *ca.* -20 cm$^{-1}$ at 5 Å, while $L_1$ has a repulsive potential curve. The stability of the configurations depends indeed on the steric hindrance: this effect is evident in the L configurations. Intermolecular forces such as dipole – dipole interactions and dispersion forces might also play a role, especially in the higher stability of $L_3$ with respect $L_2$ and $L_1$. The X and H configurations are those for which the equilibrium distance is the minimum. For the L configurations, the linear mutual orientation determines a high equilibrium distance, or even the absence of a potential well. The most stable configurations are those of the Z and T series. The lowest rms (root mean square) error among the leading configurations is 4.61 x 10$^{-4}$ cm$^{-1}$ for the X configuration, while the highest one is 1.33 cm$^{-1}$ for the $L_1$ configuration.

## 5.2. The CO – HF case.

Figure 5 (as done in Figure 4, the plots are divided into four panels) shows interesting results related to the CO – HF system: despite we reported for the CO – CO system, $L_3$ and $L_4$ are the most stable configurations. $L_3$ has a minimum energy of *ca.* -1200 cm$^{-1}$ at 3.1 Å, while $L_4$ presents an equilibrium energy of *ca.* -600 cm$^{-1}$ at the same equilibrium distance of $L_3$. The $L_1$ configuration has a very small minimum energy, while $L_2$ is repulsive. $T_4$ is the only T configuration with a significant attractive component, while for $T_2$ and $T_3$ the well depth is lower than 50 cm$^{-1}$. Regarding the Z configurations, $Z_1$ is the only stable, being its equilibrium energy  *ca.* -100 cm$^{-1}$ at 4 Å. Both $H_1$ and $H_2$ present a minimum around 3 Å: -150 cm$^{-1}$ for $H_2$ and -50 cm$^{-1}$ for $H_1$. Finally, the X configuration presents a behavior similar to that of $H_2$.

Figure 6 reports the isotropic components of the interaction potential of CO – CO and CO – HF systems. These components, which can be measured experimentally, would allow comparison and evaluation of the reliability of the theoretical method. The isotropic component of the potential, indicated as $v_{000}$, corresponds to the full averaging over the angles $\phi$, $\theta_1$, $\theta_2$. For the CO – CO system, in Figure 6 (a), the minimum is located at *ca* 4.5 Å of distance between the centers of mass of the molecules and at -50 cm$^{-1}$; $v_{000}$ is the only component with a minimum, while the other components, the anisotropic ones, have a sole repulsive character. Regarding the CO – HF system, in Figure 6 (b), the isotropic component has a minimum at *ca.* 4.5 Å and -60 cm$^{-1}$, also some anisotropic components present an attractive contribution like $v_{022}$, $v_{202}$ and $v_{224}$, whose minima are placed at 4.0 Å, 4.2 Å, 3.0 Å and -8 cm$^{-1}$, -10 cm$^{-1}$ and -50 cm$^{-1}$, respectively.

In Figure 7, we show a cut of the potential energy surface of the CO – CO system, corresponding to R= 3.9 Å, $\phi$ = 180°, $\theta_1$ =106.24° and $\theta_2$ = 40.64°, at an energy value of -126.9 cm$^{-1}$. Figure 8 shows, similarly, a cut of the potential energy surface for CO – HF, around the variables of R = 3.1 Å, $\phi$ = 157.5°, $\theta_1$ = 114.5° and $\theta_2$ = 154.8°, at -1808 cm$^{-1}$. For the CO – CO system, the global minimum, at -135 cm$^{-1}$ correspond to the following configuration: R= 4.2 Å, $\phi$ = 6.5°, $\theta_1$ =8.7° and $\theta_2$ =108.9°; for the CO – HF system, the global minimum is at -1919 cm$^{-1}$, with values of R= 2.9 Å, $\phi$ = 0.0°, $\theta_1$ =155.0° and $\theta_2$ =125.6°. For CO – HF, the absence of a minimum in the interaction potential of $Z_2$, $Z_3$, $T_3$ and $L_2$, determine fittings with a high rms error: 1.55, 43.76, 14.38 and 18.98 cm$^{-1}$, respectively. The lowest rms error, 1.35 x 10$^{-3}$ cm$^{-1}$, is given by the $T_1$ configuration.

In Figure 9, we report the comparison between our method (spherical expansion) and the single point energy calculated *ab initio* for non-leading configurations, as shown in References [21, 22]. For the CO – CO system (Figure 9a), the comparison made for the configuration with $\theta_1$ = 15.0° and $\theta_2$ =110.0° $\phi$ = 10°, shows a good agreement between the two methods. For the CO – HF (Figure 9b), configuration with $\theta_1$ = 114.5° and $\theta_2$ =154.8° $\phi$ = 157.5°, the spherical expansion overestimates the minimum (absolute value) calculate by *ab initio* method, while they show a good agreement at long range.

## 5.3. Molecular properties by quantum mechanical calculations.

For sake of completeness, we report the molecular properties of CO and HF calculated *ab* initio. Table 1 shows comparison of geometry, frequency, electric properties and energy of the CO and HF molecules calculated by using different basis sets at CCSD (T) level of theory with experimental reference data. For the HF bond-length the smallest error is obtained with cc-pV5Z basis set with $4.9 \cdot 10^{-5}$ Å, followed by cc-pVTZ with $4.6 \cdot 10^{-4}$ Å, for the frequency the best agreement is obtained by CBS with 3.8 cm$^{-1}$ and aug-cc-pV5Z with 4.2 cm$^{-1}$. Regarding the electric properties, the aug-cc-pV5Z basis set gives errors of 0.0830 D and 0.0768 Å$^3$ for the permanent dipole moment and polarizability respectively, while the aug-cc-pVQZ basis set gives 0.0841 D and 0.1046 Å$^3$. For the CO molecule the most accurate result is given by the ccpV5Z basis set with an error of $2.5 \cdot 10^{-3}$ Å and by the aug-cc-pV5Z basis set $2.7 \cdot 10^{-4}$ Å, regarding the frequency cc-pV5Z gives an error of 4.4 cm$^{-1}$ and the at cc-pVQZ basis set with 5.3 cm$^{-1}$. For the electric properties, cc-pV5Z reports an error of 0.008 D and 0.003 Å$^3$ for the dipole moment and polarizability respectively, while cc-pVQZ gives an error of 0.012 D and 0.004 Å$^3$. On the basis of the accuracy and the computational cost, we concluded that the aug-cc-pVQZ basis offers the best compromise to perform calculations on the potential energy surface points.

In Figures 10 and 11, we report the potential energy profiles of the leading configurations for the CO···CO and CO···HF systems. The *ab initio* points are fitted by the Generalized Rydberg function:

$$V(R) = -D_e \left( 1 + \sum_{i=1}^{n} a_i \left( R - R_{eq} \right)^i \right) e^{\left( -a_1 \left( R - R_{eq} \right) \right)} + E_{ref} \qquad (14)$$

where $R_{eq}$ is the equilibrium distance, $D_e$ the dissociation energy, the terms $a_i$ are adjustable parameters and $E_{ref}$ is the reference energy (see Table 2 for the adjustable parameters of the CO – CO system and Table 3 for the CO – HF system).

Table 4 reports the isotropic components of the interaction potential for CO – CO and CO – HF system, calculated by correlation formulas with and without induction correction for various basis sets, as shown in Reference [47]. As well as the isotropic component of the potential, the anisotropic components as a function of the distance (Figure 12 and 13) permit direct comparison with experimental results (see for example [48, 49]. Finally, Table 5 reports the fitting parameters for the isotropic term ($v_{000}$) of both CO – CO and CO – HF systems. The rms are 0.012 cm$^{-1}$ for CO – CO and 0.47 cm$^{-1}$ for CO –HF.

# 6. Conclusions.

The method we have reported was developed to characterize the spectroscopy of weakly bound aggregates and of collisional processes. On this purpose, the main target is to give an exact transformation of the input data that corresponds to those of a minimal set of configurations, the *leading configurations*, which are those relevant in the process we consider. The procedure permits both the interpolation and extrapolation to structures beyond those corresponding to the leading configurations, and as a key point, expansibility to a larger number of minimal configurations and a more accurate level of theory, when available.

This work completes the study of potential energy surfaces, obtained by combining quantum chemical calculations and spherical harmonics expansions of van der Waals' clusters composed by couples of diatomic molecules. This account included a revisitation of the simplest cases $N_2$-$N_2$, $O_2$-$O_2$ and $N_2$-$O_2$, including applications to the validation of experimental results and demonstrating the reliability of the method. The main target of the paper is to cover the generality of possible cases, providing the explicit extension of the method to the more complex and less symmetric systems, namely those involving either a homonuclear diatomic molecule interacting with a heteronuclear one, or the interaction of both two identical and different heteronuclear molecules. As applications of the method, we have presented the spherical harmonic expansions for the quantum mechanically generated potential energy surfaces for CO – CO and CO – HF.

## Acknowledgment.

We are pleased to dedicate this article to the special issue of Journal of Molecular Spectroscopy focusing on ''Spectroscopy and Inter/Intramolecular Dynamics in Honor of Walther Caminati". F. P., A. L. and V. A. acknowledges the Italian Ministry for Education, University and Research, MIUR, for financial supporting: SIR 2014 "Scientific Independence for young Researchers" (RBSI14U3VF). Vincenzo Aquilanti thanks CAPES for the appointment as Professor Visitante Especial at Instituto de Fisica, Universidade Federal de Bahia, Salvador (Brazil).

# References.

**Figures captions.**

**Figure 1.** We report the coordinate reference system for CO – CO (the description can be easily extended to the general case AB – CD). The CO – CO system is embedded in the *xyz* Cartesian reference frame, where the origin of the axes coincides with the center-of-mass of the system. Two additional axes *z'* and *z''*, parallel to the C – O bonds are defined. The system is described by four variables: the distance between the centers-of-mass of the molecules, *R*; the polar angles between the *z'*- and the *z*-axis, $\theta_1$, and the *z''*- and the *z*-axis $\theta_2$ ($0 \leq \theta_1$, $\theta_2 \leq \pi$); the dihedral angle $\phi$ ($0 \leq \phi < 2\pi$). The bond distances $r_1$ and $r_2$ are considered constant.

**Figure 2.** The eleven leading configurations of the CO – CO system. Within parenthesis we report the three angles $\theta_1$, $\theta_2$ and $\phi$.

**Figure 3.** The fourteen leading configurations for the CO – HF system. Within parenthesis we reported the three angles $\theta_1$, $\theta_2$ and $\phi$.

**Figure 4.** CO – CO interaction energies for the leading configurations reported in Figure 2 as a function of the reciprocal distances of the centers of mass of the two molecules.

**Figure 5.** CO – HF interaction energies for the leading configurations in Figure 3 as a function of the reciprocal distances of the centers of mass of the two molecules.

**Figure 6.** Dependence on molecule – molecule distance of the isotropic moments, $v_{000}$ $(R)$ of the spherical expansion for CO – CO (a) and CO – HF (b) systems.

**Figure 7.** Representation of the potential energy surface of the CO – CO system near the global minimum configuration, corresponding to R= 3.8 Å, $\phi = 0°$, $\theta_1 =106.24°$ and $\theta_2 =40.64°$, at the energy of -126.9 cm$^{-1}$.

**Figure 8.** Representation of the potential energy surface of the CO – HF system near the global minimum configuration, corresponding to R = 3.1 Å, $\phi = 157.5°$, $\theta_1 = 114.5°$ and $\theta_2 = 154.8°$, at the energy of -1808 cm$^{-1}$.

**Figure 9.** The figure shows a comparison between the intermolecular potential calculated *ab initio* (circles) and by the spherical expansion (continuous line) for non-leading configurations of the CO – CO system ($\theta_1=15°$, $\theta_2=110°$, $\phi=10°$) and CO – HF system ($\theta_1=114.5°$, $\theta_2=154.8°$, $\phi=157.5°$).

**Figure 10.** The most relevant *ab initio* points of the eleven leading configurations of the CO – CO system fitted by the Generalized Rydberg function.

**Figure 11.** The most relevant *ab initio* points of the fourteen leading configurations of the CO – HF system fitted by the Generalized Rydberg function.

**Figure 12.** Anisotropic components of the interaction potential (in cm$^{-1}$) as a function of the distance (in Å) of the CO – CO system.

**Figure 13.** Anisotropic components of the interaction potential (in cm$^{-1}$) as a function of the distance (in Å) of the CO – HF system.

**Figure1.**

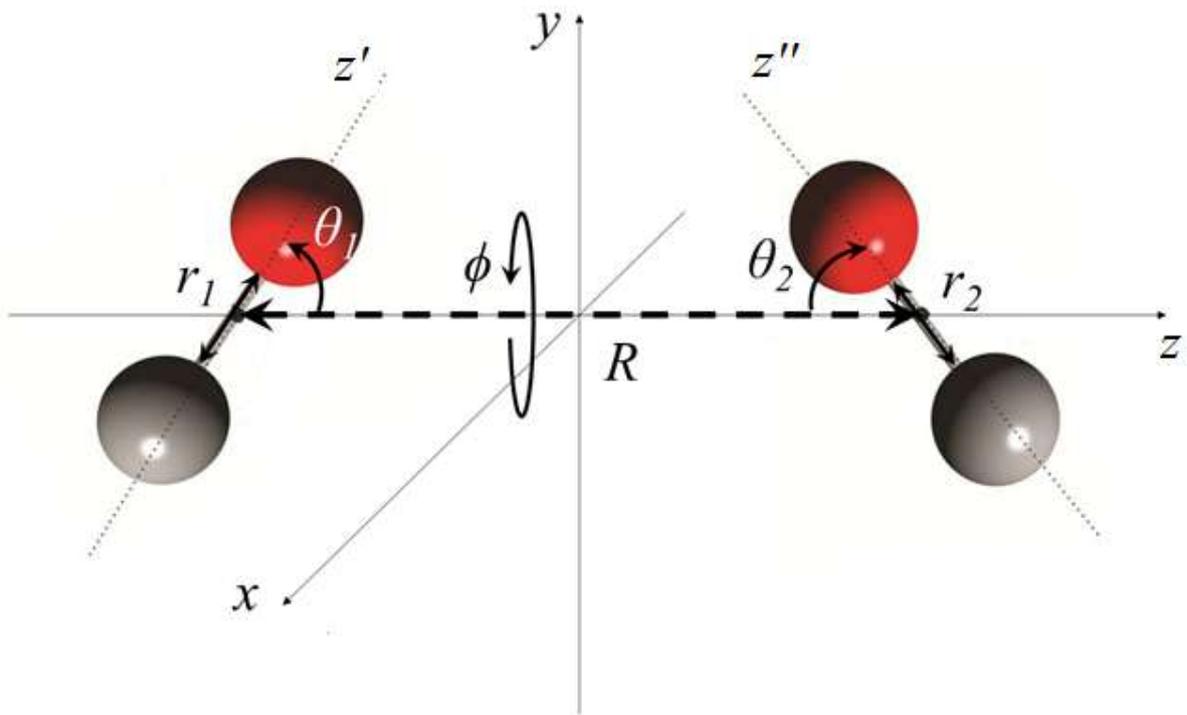

**Figure 2.**

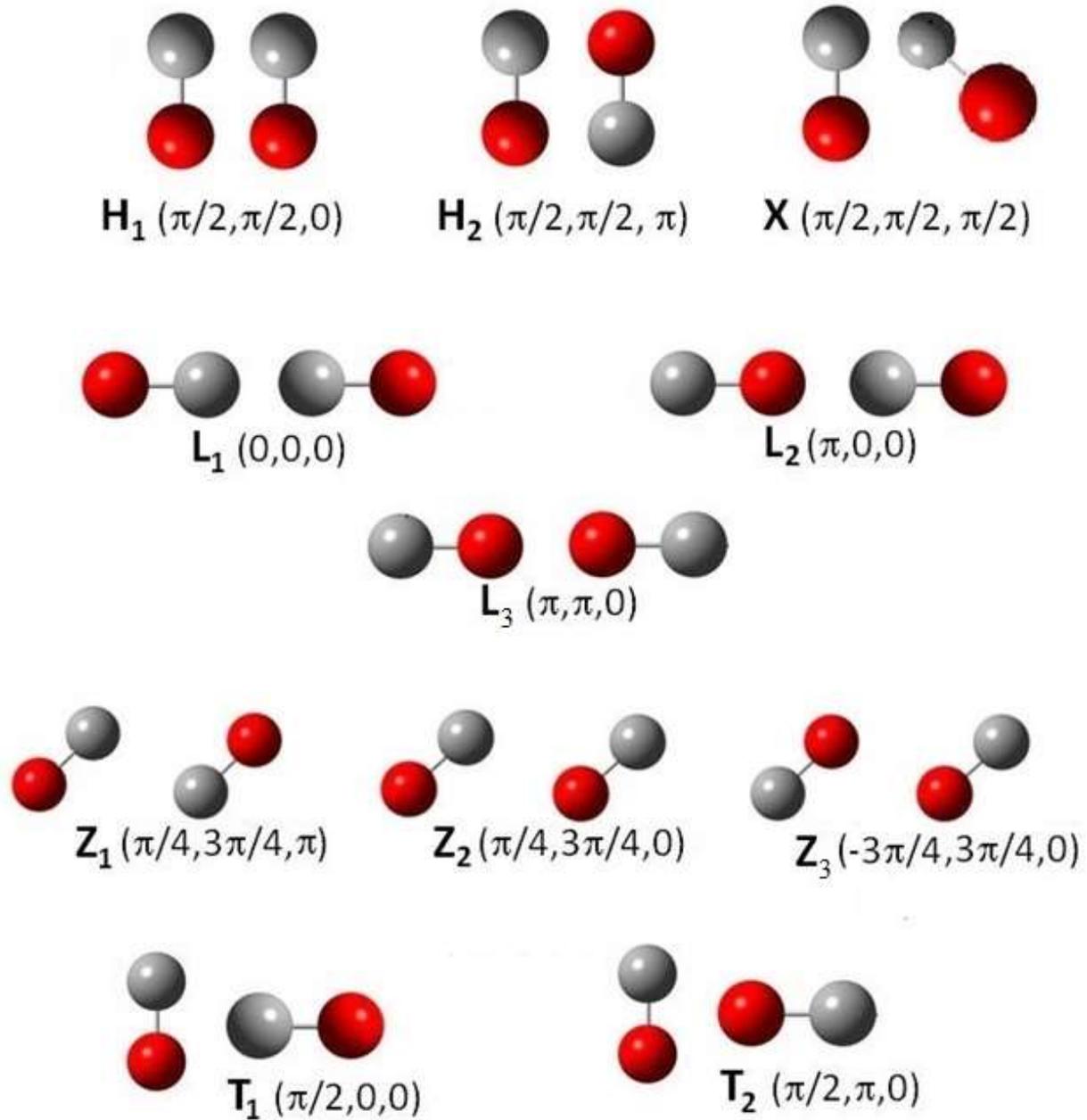

Leading Configurations CO - CO

$H_1$ $(\pi/2, \pi/2, 0)$    $H_2$ $(\pi/2, \pi/2, \pi)$    $X$ $(\pi/2, \pi/2, \pi/2)$

$L_1$ $(0,0,0)$    $L_2$ $(\pi, 0, 0)$

$L_3$ $(\pi, \pi, 0)$

$Z_1$ $(\pi/4, 3\pi/4, \pi)$    $Z_2$ $(\pi/4, 3\pi/4, 0)$    $Z_3$ $(-3\pi/4, 3\pi/4, 0)$

$T_1$ $(\pi/2, 0, 0)$    $T_2$ $(\pi/2, \pi, 0)$

**Figure 3.**

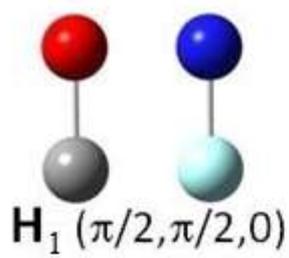 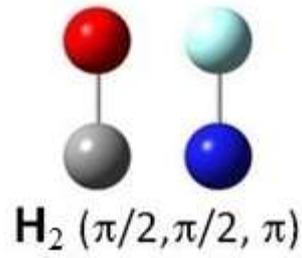 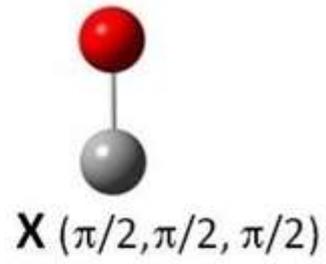

**H**₁ ($\pi/2, \pi/2, 0$)    **H**₂ ($\pi/2, \pi/2, \pi$)    **X** ($\pi/2, \pi/2, \pi/2$)

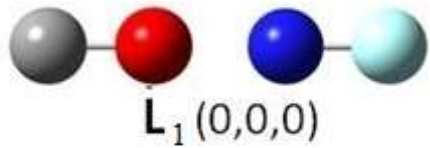 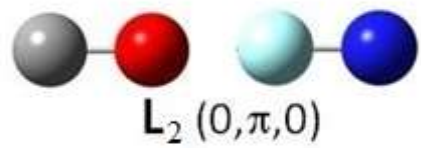

**L**₁ ($0,0,0$)    **L**₂ ($0, \pi, 0$)

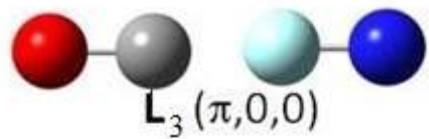 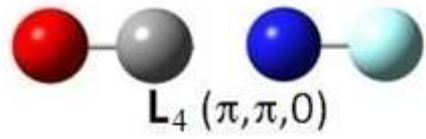

**L**₃ ($\pi, 0, 0$)    **L**₄ ($\pi, \pi, 0$)

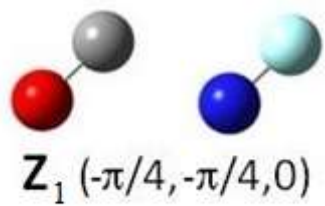 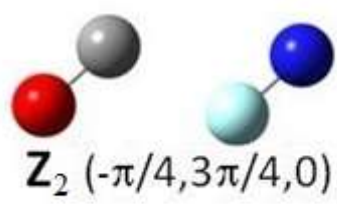 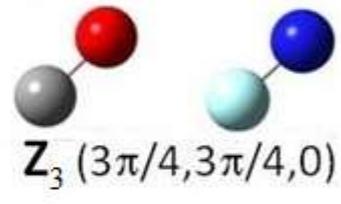

**Z**₁ ($-\pi/4, -\pi/4, 0$)    **Z**₂ ($-\pi/4, 3\pi/4, 0$)    **Z**₃ ($3\pi/4, 3\pi/4, 0$)

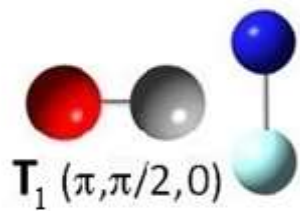 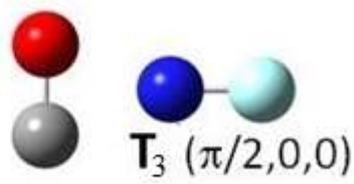

**T**₁ ($\pi, \pi/2, 0$)    **T**₃ ($\pi/2, 0, 0$)

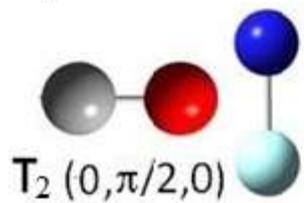 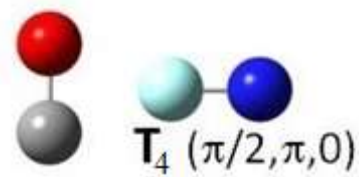

**T**₂ ($0, \pi/2, 0$)    **T**₄ ($\pi/2, \pi, 0$)

**Figure 4.**

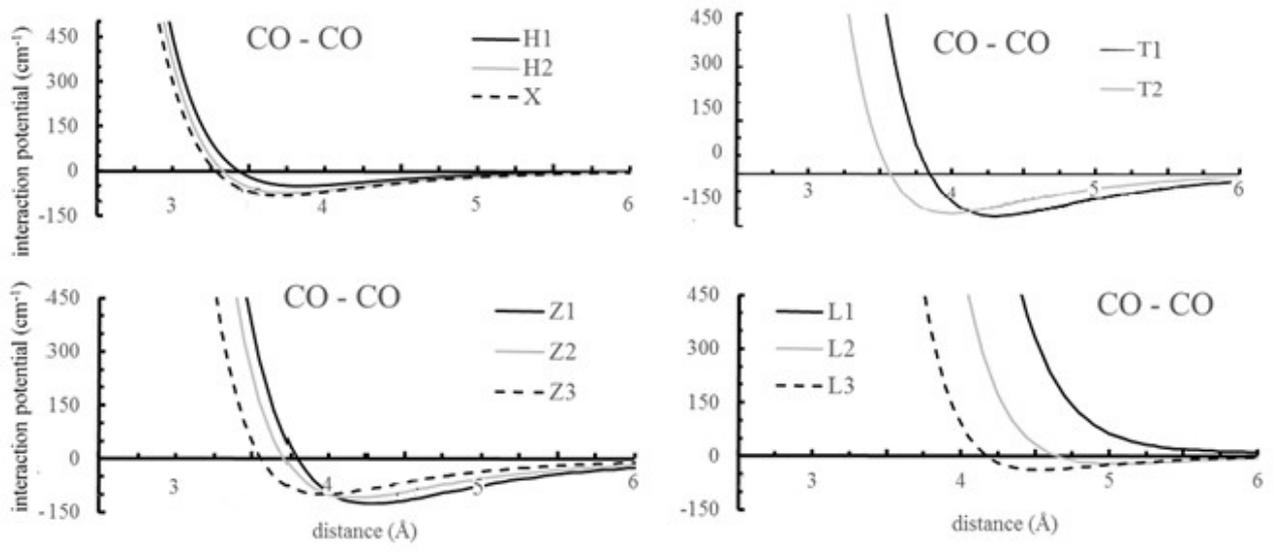

**Figure 5.**

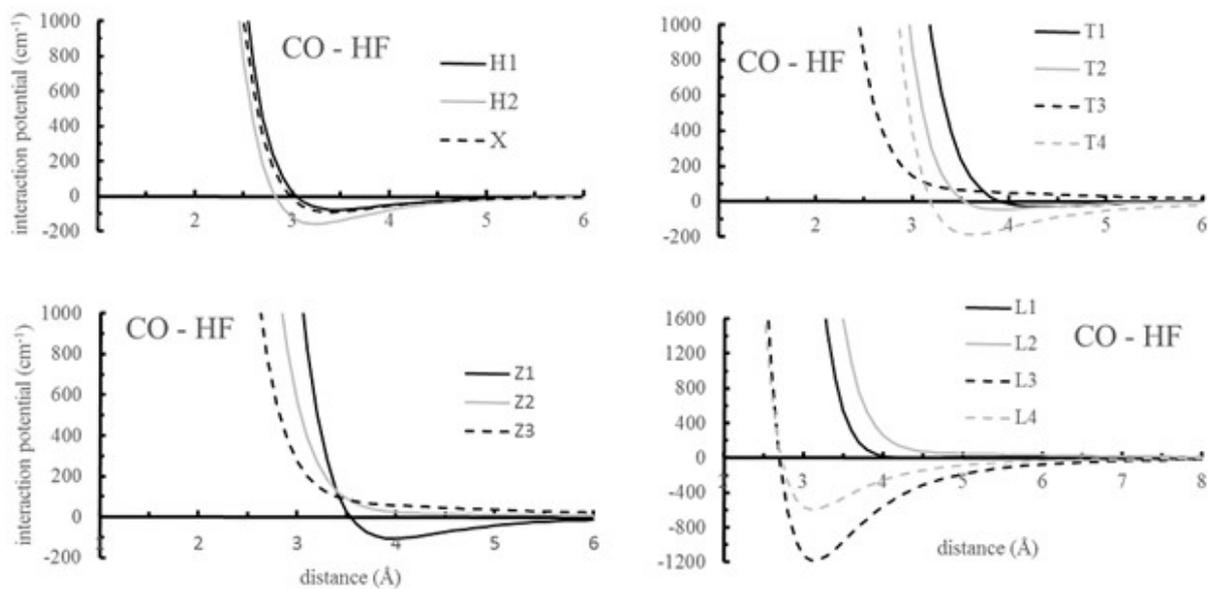

**Figure 6.**

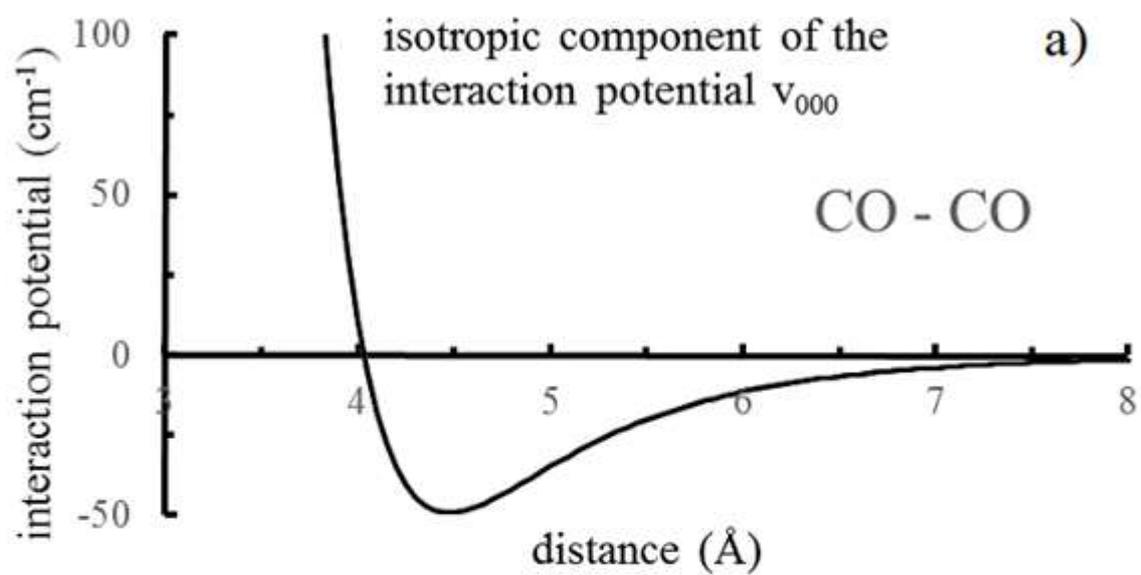

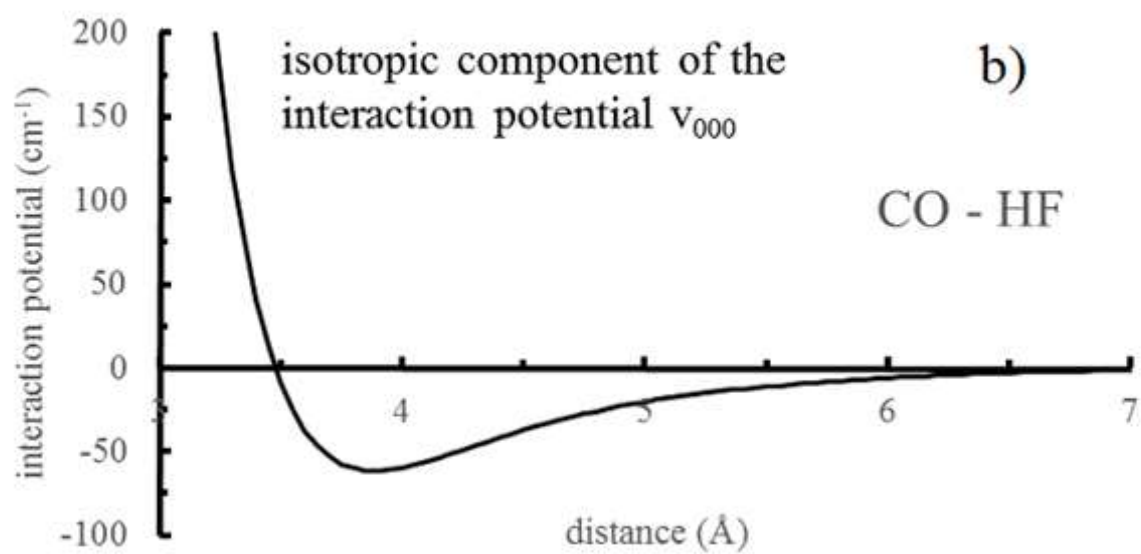

**Figure 7.**

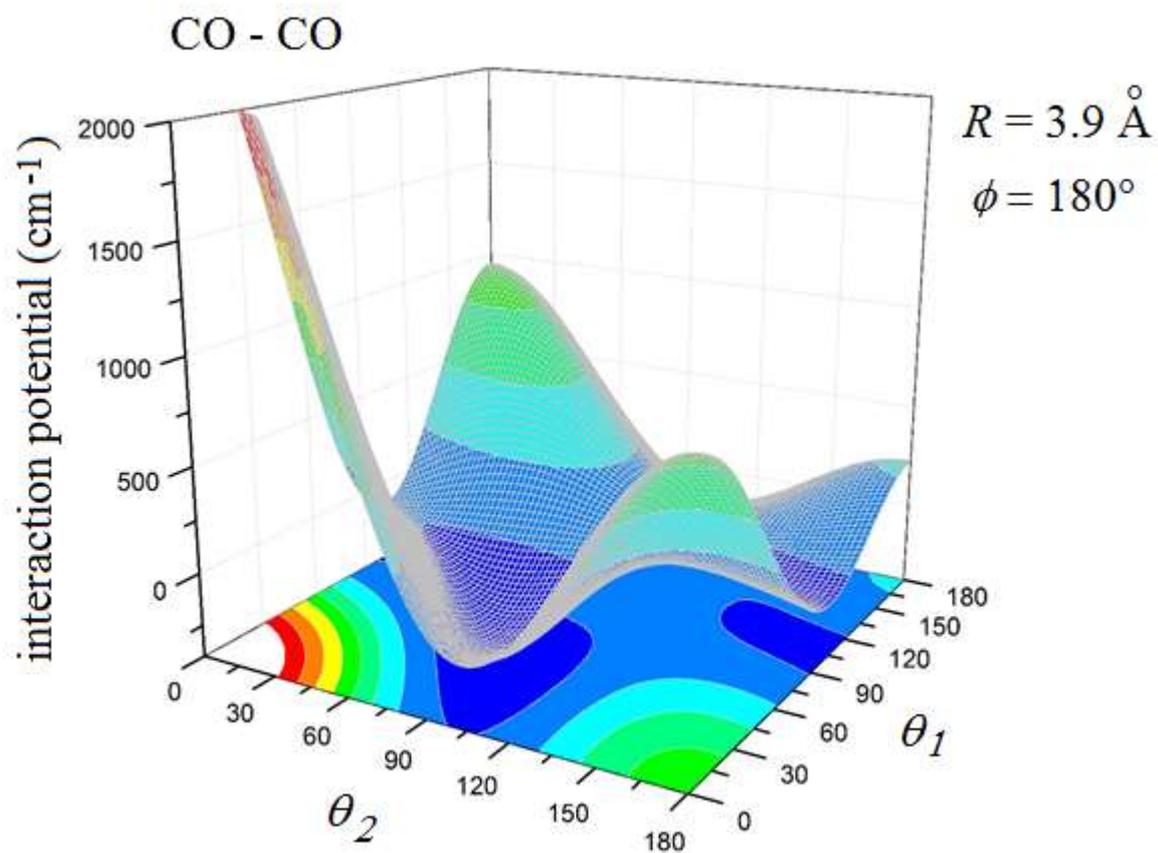

**Figure 8.**

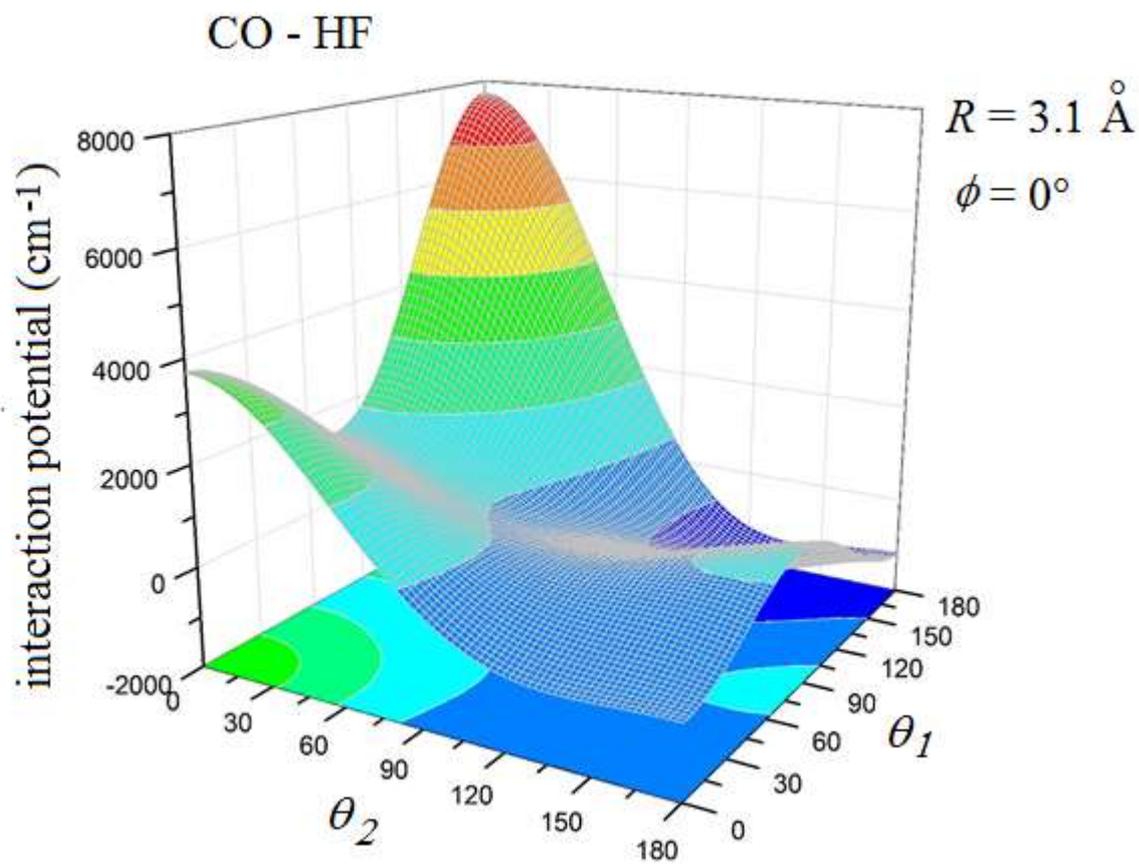

**Figure 9.**

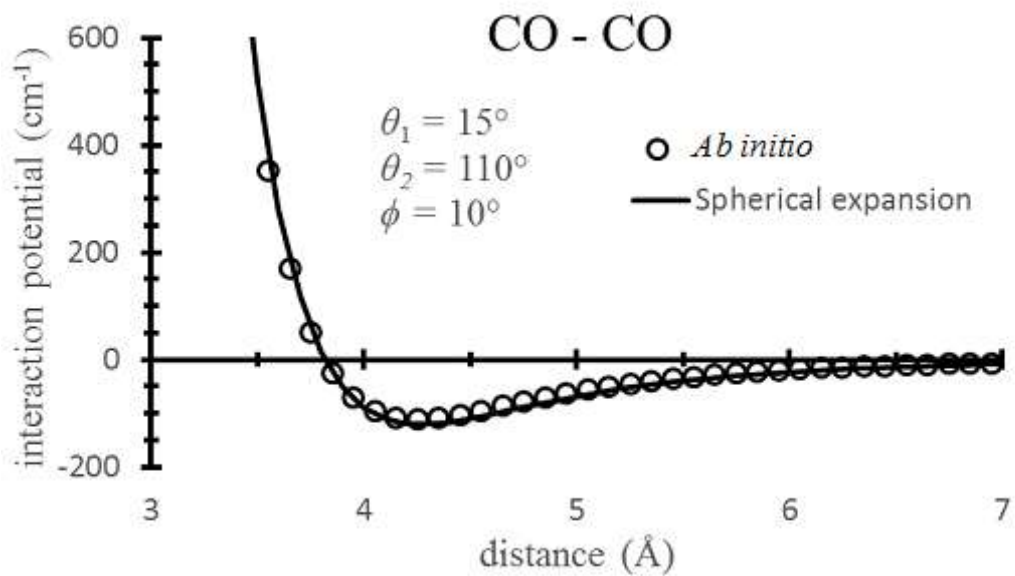

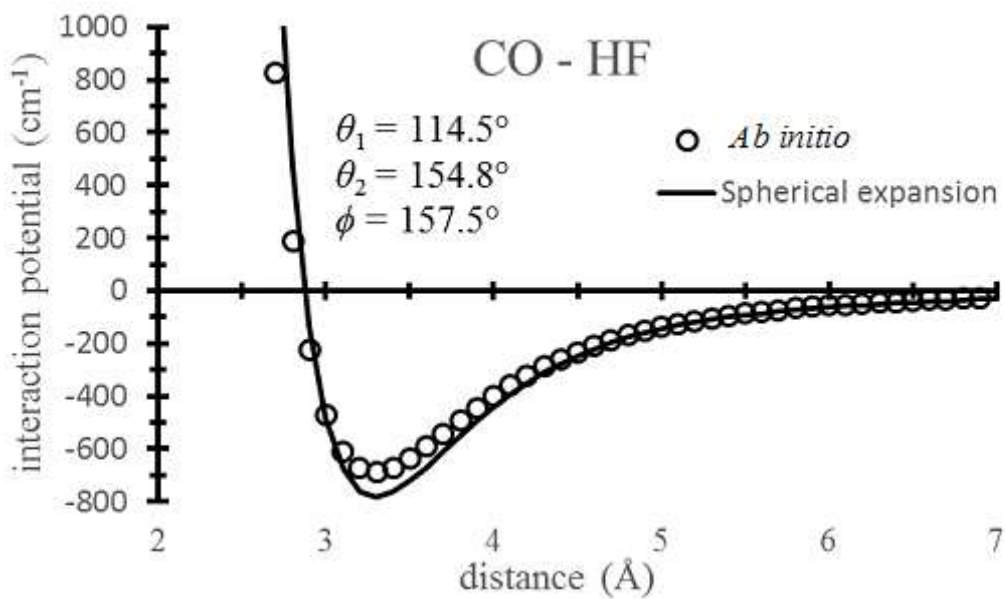

**Figure 10.**

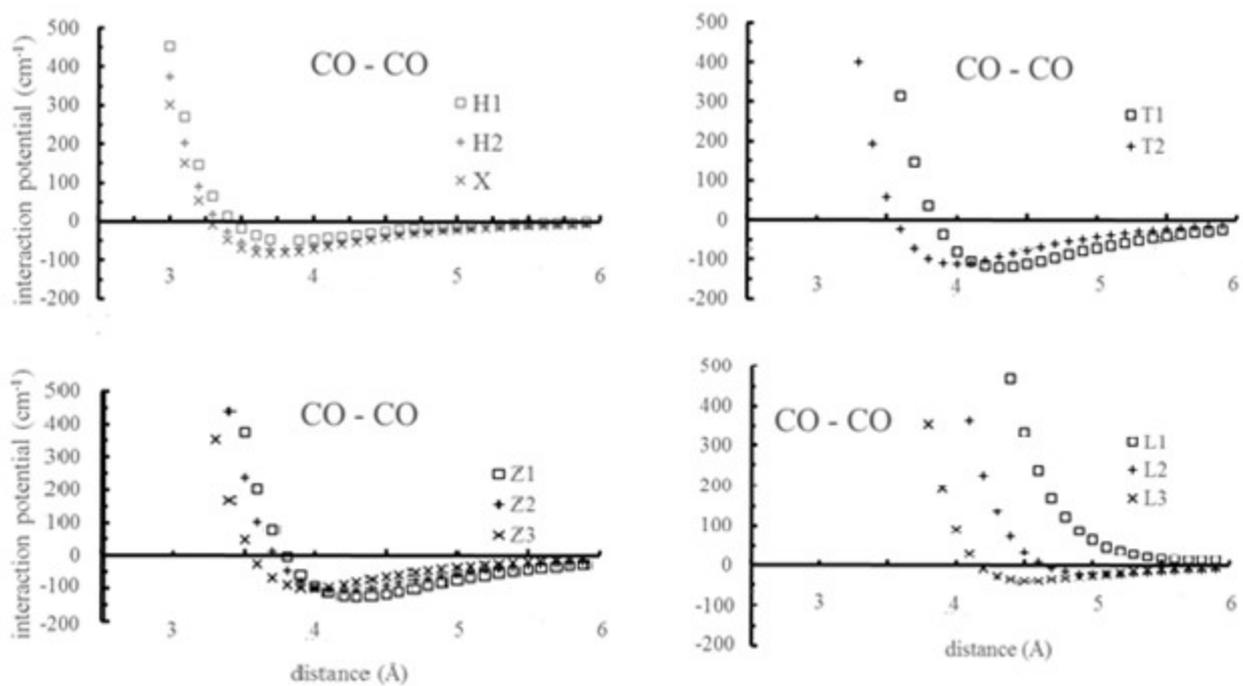

**Figure 11.**

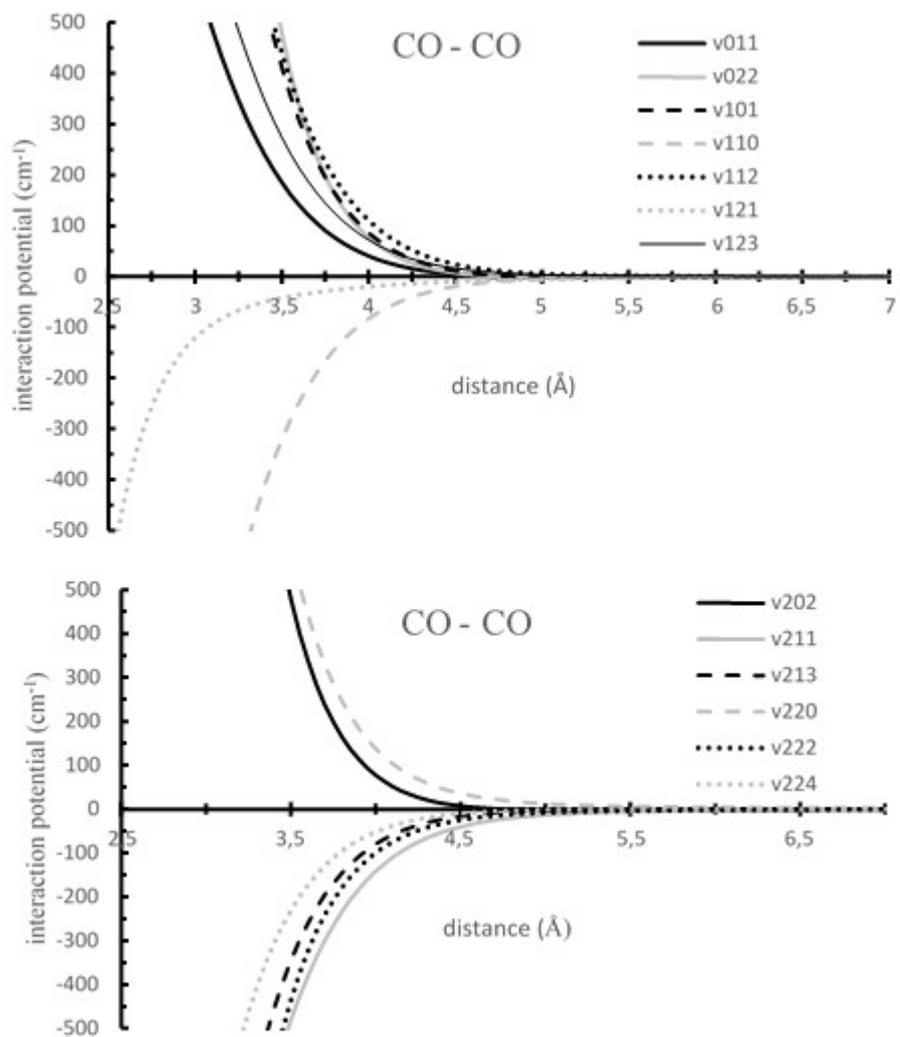

**Figure 12.**

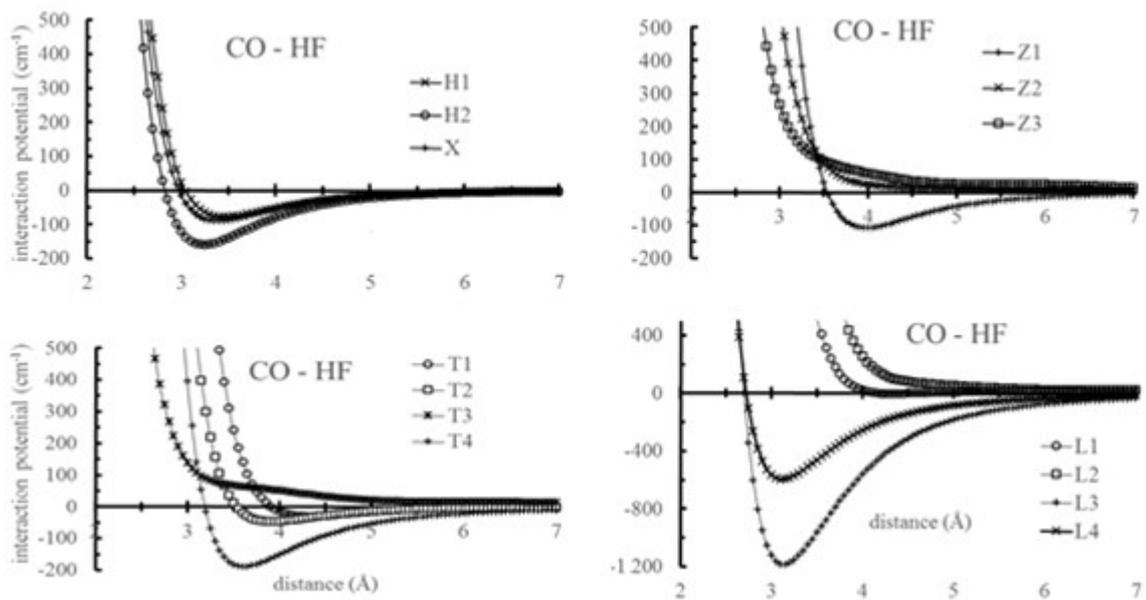

**Figure 13.**

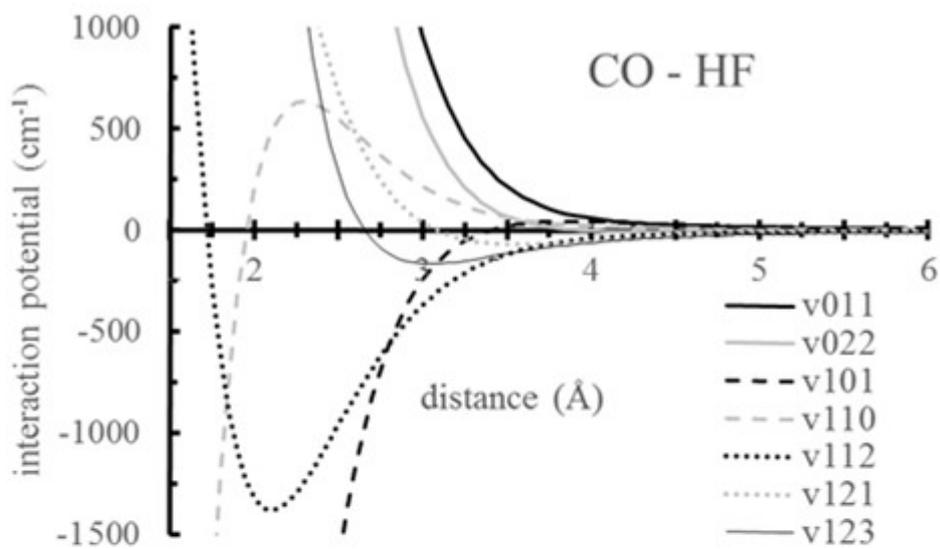

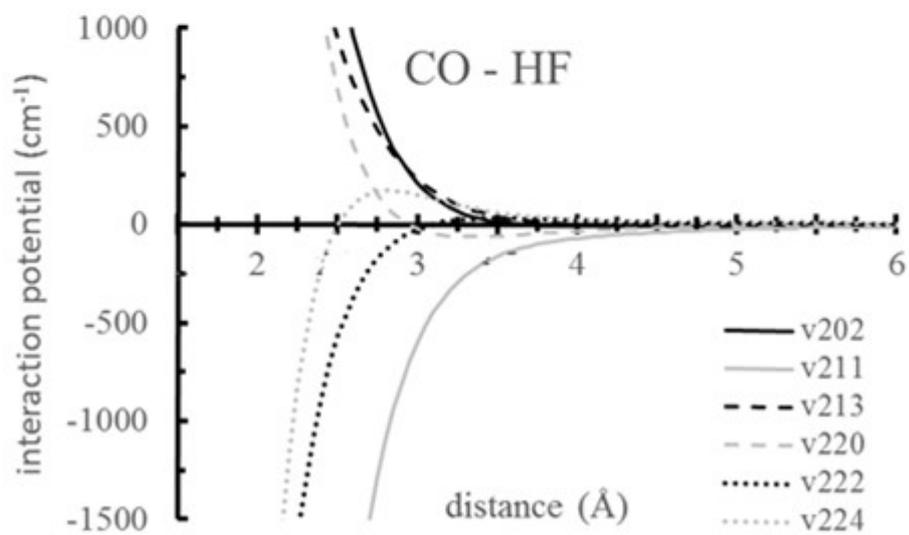

**Table 1.** Geometrical parameters and electronic properties of CO and HF molecules obtained from various basis sets and Complete Basis Sets extrapolation, compared with reference experimental data. $R_{eq}$ (in Å) is the equilibrium distance, $\omega$ (in cm$^{-1}$) is the stretching frequency, $\mu$ (in Debye) is the permanent dipole moment, $\alpha$ (in Å$^3$) is the polarizability and E (in hartree) is the equilibrium energy.

| HF | | | | | |
|---|---|---|---|---|---|
| | $R_{eq}$ (Å) | $\omega$ (cm$^{-1}$) | $\mu$(D) | $\alpha$(Å$^3$) | E(hartree) |
| cc-pVDZ | 0.920 | 4151.0 | 1.945 | 2.468 | -100.228 |
| aug-cc-pVDZ | 0.924 | 4081.4 | 1.931 | 4.955 | -100.264 |
| cc-pVTZ | 0.917 | 4177.9 | 1.929 | 3.554 | -100.338 |
| aug-cc-pVTZ | 0.921 | 4125.5 | 1.919 | 5.377 | -100.350 |
| cc-pVQZ | 0.916 | 4163.1 | 1.919 | 4.230 | -100.703 |
| aug-cc-pVQZ | 0.918 | 4142.8 | 1.909 | 5.495 | -100.377 |
| cc-pV5Z | 0.917 | 4151.8 | 1.918 | 4.694 | -100.385 |
| aug-cc-pV5Z | 0.917 | 4142.6 | 1.908 | 5.523 | -100.387 |
| CBS [40] | 0.918 | 4142.2 | 1.917 | 4.820 | -100.418 |
| Reference | 0.917 [41] | 4138.4 [42] | 1.825 [43] | 5.600 [44] | |

| CO | | | | | |
|---|---|---|---|---|---|
| | $R_{eq}$ (Å) | $\Omega$ (cm$^{-1}$) | $\mu$ (D) | $A$ (Å$^3$) | $E$ (hartree) |
| cc-pVDZ | 1.145 | 2143.4 | 0.171 | 9.779 | -113.055 |
| aug-cc-pVDZ | 1.147 | 2104.8 | 0.080 | 13.155 | -113.074 |
| cc-pVTZ | 1.136 | 2153.6 | 0.143 | 11.399 | -113.156 |
| aug-cc-pVTZ | 1.136 | 2144.8 | 0.101 | 13.166 | -113.162 |
| cc-pVQZ | 1.131 | 2164.7 | 0.115 | 12.160 | -113.188 |
| aug-cc-pVQZ | 1.131 | 2160.1 | 0.105 | 13.186 | -113.190 |
| cc-pV5Z | 1.131 | 2165.6 | 0.108 | 12.713 | -113.198 |
| aug-cc-pV5Z | 1.131 | 2164.0 | 0.106 | 13.137 | -113.199 |
| CBS [40] | 1.133 | 2153.4 | 0.107 | 12.839 | -113.176 |
| Reference | 1.128 [41] | 2170.0 [44] | 0.11 [43] | 13.178 [45] | |

**Table 2.** Rydberg fitting parameters for the eleven leading configurations of the CO – CO system.

| | $a_1$ | $a_2$ | $a_3$ | $a_4$ | $a_5$ | $D_e$ | $R_{eq}$ | $E_{ref}$ | rms |
|---|---|---|---|---|---|---|---|---|---|
| $H_1$ | 1.9727 | -0.9889 | 0.5133 | -0.1031 | -1.3680E-09 | 50.1993 | 3.8414 | 0.1170 | $4.88 \cdot 10^{-3}$ |
| $H_2$ | 1.8040 | -1.0851 | 0.6837 | -0.1779 | 2.9189E-02 | 73.1301 | 3.7504 | 2.7691E-02 | $1.48 \cdot 10^{-2}$ |
| X | 1.7669 | -1.0062 | 0.6232 | -0.1465 | 2.2705E-02 | 83.5949 | 3.7104 | -2.0291E-02 | $4.61 \cdot 10^{-4}$ |
| $Z_1$ | 1.9908 | -0.0374 | 0.1698 | 0.1171 | 2.3155E-09 | 125.8989 | 4.2849 | -0.4009 | $7.33 \cdot 10^{-2}$ |
| $Z_2$ | 1.5504 | -1.0831 | 0.6882 | -0.1714 | 2.2828E-02 | 109.8599 | 4.1690 | 0.1154 | $6.65 \cdot 10^{-3}$ |
| $Z_3$ | 1.8651 | -1.0509 | 0.8538 | -0.2382 | 4.6009E-02 | 100.1906 | 3.9669 | -5.9514E-02 | $2.45 \cdot 10^{-3}$ |
| $T_1$ | 1.6137 | -0.9053 | 0.6115 | -0.1483 | 2.3066E-02 | 121.0629 | 4.3023 | 0.4519 | $3.26 \cdot 10^{-3}$ |
| $T_2$ | 1.8865 | -0.8441 | 0.7390 | -0.1955 | 4.5261E-02 | 113.3353 | 3.9837 | 3.4653E-02 | $2.23 \cdot 10^{-3}$ |
| $L_1$ | 1.2583 | -0.3184 | 4.9122 | 4.9471 | 1.7445 | 0.3362 | 6.9036 | 7.7215 | 1.33 |
| $L_2$ | 1.9128 | -1.2363 | 0.7958 | -0.2472 | 4.3567E-02 | 22.0296 | 5.0357 | -2.1271E-02 | $5.69 \cdot 10^{-4}$ |
| $L_3$ | 2.3078 | -1.0526 | 0.8664 | -0.2461 | 2.8767E-09 | 39.5980 | 4.5304 | 0.1955 | $8.29 \cdot 10^{-3}$ |

**Table 3.** Rydberg fitting parameters for the fourteen leading configurations of the CO – HF system.

| | $a_1$ | $a_2$ | $a_3$ | $a_4$ | $a_5$ | $D_e$ | $R_{eq}$ | $E_{ref}$ | rms |
|---|---|---|---|---|---|---|---|---|---|
| $H_1$ | 1.9847 | -0.7352 | 0.6940 | -0.1819 | 0.0693 | 77.0722 | 3.4551 | -0.1869 | $1.36 \cdot 10^{-2}$ |
| $H_2$ | 1.6598 | -1.3815 | 0.8913 | -0.2514 | 0.0309 | 159.2944 | 3.2394 | 0.3093 | $8.01 \cdot 10^{-2}$ |
| $X$ | 1.9470 | -0.8787 | 0.6417 | -0.1485 | 0.0358 | 90.9793 | 3.3864 | -0.0614 | $2.41 \cdot 10^{-3}$ |
| $Z_1$ | 1.3054 | -1.5736 | 1.1447 | -0.3483 | 0.0430 | 111.3322 | 3.9764 | 3.1461 | 0.4606 |
| $Z_2$ | 1.2283 | 0.7663 | 52.8693 | 42.5351 | 9.1712 | $3.29 \cdot 10^{-2}$ | 6.1936 | 5.8178 | 1.5498 |
| $Z_3$ | 1.2662 | 0.9110 | 831907.1510 | 676474.0917 | 148195.1052 | $9.66 \cdot 10^{-6}$ | 5.5037 | 24.3118 | 43.7619 |
| $T_1$ | 2.5202 | 0.1629 | 0.3699 | 0.0978 | 0.0000 | 29.6949 | 4.3047 | -1.97E-02 | $1.35 \cdot 10^{-3}$ |
| $T_2$ | 2.2774 | -0.6947 | 0.7141 | -0.1366 | 0.0588 | 46.6464 | 3.9104 | -4.84E-02 | $2.38 \cdot 10^{-3}$ |
| $T_3$ | 1.1479 | 0.6537 | 16509.3162 | 11527.7119 | 2089.5199 | 2.62E-04 | 5.9195 | 16.2468 | 14.3752 |
| $T_4$ | 1.8724 | -0.7949 | 0.8894 | -0.2752 | 0.0765 | 188.7801 | 3.6075 | -0.3442 | $4.43 \cdot 10^{-2}$ |
| $L_1$ | 1.8400 | -29.8590 | 16.0970 | -7.2253 | 0.0000 | 3.6732 | 4.3122 | 0.5478 | 0.1021 |
| $L_2$ | 1.1978 | 0.8214 | 147744.1805 | 117519.9630 | 24853.4943 | 4.02E-05 | 6.6993 | 22.4114 | 18.9756 |
| $L_3$ | 1.8989 | -0.6763 | 0.5719 | -0.1470 | 0.0525 | 1183.9710 | 3.1329 | -2.5239 | 3.5130 |
| $L_4$ | 2.1946 | -0.5619 | 0.7855 | -0.2056 | 0.1253 | 593.0132 | 3.1164 | -2.1360 | 2.6442 |

**Table 4.** Minimum distance $R_m$ (in Å) and well depth $E$ (in cm$^{-1}$) of the isotropic components of the potential energy surface for CO···CO and CO···HF systems calculated by correlation formulas with and without induction contribution, calculated for various basis sets, adz (aug-cc-pVDZ), atz (aug-cc-pVTZ), aqz (aug-cc-pVQZ) and a5z (aug-cc-pV5Z) as shown in Reference [46]; with the Complete Basis Sets (CBS) extrapolation and by using experimental data (exp), and compared with *ab initio* calculation at CCSD(T)/aug-cc-pVQZ level of theory.

| | without induction contribution [46] | | with induction contribution [46] | | CCSD(T)/aug-cc-pVQZ | |
|---|---|---|---|---|---|---|
| | $R_m$ (Å) | $E$ (cm$^{-1}$) | $R_m$ (Å) | $E$ (cm$^{-1}$) | $R_m$ (Å) | $E$ (cm$^{-1}$) |
| CO – CO (adz) | 3.889 | 98.317 | 3.889 | 98.327 | | |
| CO – CO (atz) | 3.889 | 98.369 | 3.889 | 98.385 | | |
| CO – CO (aqz) | 3.890 | 98.465 | 3.890 | 98.483 | 4.475 | 45.984 |
| CO – CO (a5z) | 3.888 | 98.228 | 3.888 | 98.246 | | |
| CO – CO (CBS) | 3.875 | 96.798 | 3.875 | 96.816 | | |
| CO – CO (exp) | 3.890 | 98.427 | 3.890 | 98.447 | | |
| CO – HF (adz) | 3.695 | 67.952 | 3.695 | 67.958 | | |
| CO – HF (atz) | 3.688 | 66.553 | 3.688 | 66.562 | | |
| CO – HF (aqz) | 3.693 | 67.319 | 3.693 | 67.329 | 3.896 | 63.114 |
| CO – HF (a5z) | 3.692 | 67.462 | 3.692 | 67.472 | | |
| CO – HF (CBS) | 3.661 | 62.637 | 3.661 | 62.647 | | |
| CO – HF (exp) | 3.675 | 63.744 | 3.675 | 63.755 | | |

**Table 5.** Rydberg fitting parameters for the isotropic term ($v_{000}$) of the CO-CO and CO – HF system.

|  | CO-CO | CO-HF |
|---|---|---|
| $a_1$ | 1.889996 | 1.700926 |
| $a_2$ | -0.57731 | -1.18492 |
| $a_3$ | 0.493315 | 0.90756 |
| $a_4$ | -0.05609 | -0.3541 |
| $a_5$ | 0.024949 | 0.079861 |
| $D_e$ | 49.5122 | 63.1143 |
| $R_{eq}$ | 4.4755 | 3.8962 |
| $E_{ref}$ | 0.2335 | 1.6217 |
| rms | 0.012 | 0.47 |